\documentclass[preprint,11pt]{aastex}

\newcommand{\lam}{$\lambda$}

\newcommand{\ecss}{erg~cm$^{-2}$~s$^{-1}$~sr$^{-1}$} % erg/cm2/s/sr
\newcommand{\ecs}{erg~cm$^{-2}$~s$^{-1}$} % erg/cm2/s
\newcommand{\kms}{km~s$^{-1}$}

\renewcommand{\ion}[2]{#1\,{\sc #2}}

\newcommand{\hinode}{\emph{Hinode}}

\def\ionx[#1 #2]{#1\,{\sc #2}}

\title{Properties of a solar flare kernel observed by Hinode and SDO}

\author{P. R. Young\altaffilmark{1},
G. A. Doschek\altaffilmark{2},
H. P. Warren\altaffilmark{2}
\and H. Hara\altaffilmark{3}
}

\altaffiltext{1}{College of Science, George Mason University, 4400 University Drive, Fairfax, VA 22030}
\altaffiltext{2}{Naval Research Laboratory, 4555 Overlook Avenue SW,
  Washington DC 20375} 
\altaffiltext{3}{National Astronomical Observatory of Japan/NINS,
  2-21-1 Osawa, Mitaka, Tokyo 181-8588, Japan}

\begin{document}

\begin{abstract}
Flare kernels are compact features located in the solar chromosphere that
are the sites of rapid heating and plasma upflow during the rise phase
of flares. 
An example is presented from a M1.1 class flare observed
on 2011 February 16 07:44~UT for which the location of the upflow region
seen by EIS can be precisely aligned to high spatial resolution images
obtained by the Atmospheric Imaging Assembly (AIA) and Heliospheric
Magnetic Imager (HMI) on board the Solar Dynamics
Observatory (SDO). A string of bright flare
kernels is found to be aligned with a ridge of strong magnetic field,
and one kernel site is highlighted for which an upflow speed of $\approx
400$~\kms\ is measured in lines formed at 10--30~MK. The line-of-sight
magnetic field strength at this location is $\approx$~1000~G.
Emission over a continuous range of temperatures down to the
chromosphere is found, and the kernels
have a similar morphology at all temperatures and are spatially
coincident with sizes at the 
resolution 
limit of the AIA instrument ($\lesssim$~400~km).
For temperatures of 0.3--3.0~MK the EIS emission lines show multiple
velocity components, with the dominant component becoming more
blue-shifted with temperature from a redshift of 35~\kms\ at 0.3~MK to
a blueshift of 60~\kms\ at 3.0~MK. Emission lines from 1.5--3.0~MK
show a weak redshifted component at around 60--70~\kms\ implying
multi-directional flows at the kernel site. Significant non-thermal
broadening corresponding to velocities of $\approx$~120~\kms\ is found
at 10--30~MK, and the electron density in
the kernel, 
measured at 2~MK, is $3.4\times 10^{10}$~cm$^{-3}$. Finally, 
the \ion{Fe}{xxiv} \lam192.03/\lam255.11 ratio suggests that the
EIS calibration has changed since launch, with the
long wavelength channel less sensitive than the short wavelength
channel by around a factor two.
\end{abstract}


\section{Introduction}

Early spectroscopic measurements of emission lines  formed at
$\sim$10~MK obtained during the rise
phase of solar flares revealed blue-shifted components corresponding to
plasma upflows of several hundred \kms\
\citep{doschek80,antonucci82}. Theoretical models of 1D solar loops in
which energy is deposited near the loop top demonstrated that such
large upflows could be generated as low-lying chromospheric plasma is
heated to coronal temperatures and ``evaporates'' into  the coronal
part of the loop \citep{1983ApJ...265.1090C}. The plasma upflow sites
have also been correlated with brightenings in the chromosphere
\citep{1986ApJ...309..435M}, transition region and corona
\citep{2006SoPh..234...95D}. In the present work we define intense brightenings
that occur during the flare rise phase as 
\emph{flare kernels}. It is not known if flare kernels always
exhibit fast upflows in $\sim$10~MK emission lines but  the many
reported measurements of hot, blue-shifted emission lines in X-ray and
ultraviolet spectra suggests that this is possbile.

The availability since 2010 of high spatial and time resolution
EUV images from the Atmospheric Imaging Assembly (AIA) on board the Solar
Dynamics Observatory (SDO), coupled with high 
resolution EUV spectra from the EUV Imaging Spectrometer (EIS) on
board \hinode\ gives an unprecedented capability for studying flare
kernels. The present work focusses on one particular kernel observed
by both instruments on 2011 February 16 during the rise phase of a M1
class flare.

Flare kernels can be interpreted in terms of the Standard Flare Model
\citep[see][and references therein]{benz08}, which posits an
energy release site in the corona that leads to either a
stream of non-thermal particles or a thermal conduction front being
directed down the coronal loop legs towards the photosphere. Heating then occurs in the chromosphere
leading to flare ribbons and brightenings, and hot plasma rises
towards the flare site, giving bright post-flare loops. The flare kernels are
then the sites of chromospheric heating, and would be expected to be
bright in emission lines formed at all temperatures from the
chromosphere to $\sim$10~MK. In addition, large plasma flows would
be expected as heated plasma rises into the corona. The case for which
non-thermal particles heat the chromosphere has been investigated
theoretically, and two evaporation scenarios identified: for low particle
fluxes ($< 3\times 10^{10}$~\ecs) ``gentle'' evaporation occurs with
speeds up to 30~\kms; while for high particle fluxes ``explosive''
evaporation occurs with speeds of several hundred \kms\ in the
corona \citep{1985ApJ...289..414F}. Evidence for both evaporation scenarios has been found
\citep[e.g.,][]{2006ApJ...642L.169M,2003ApJ...588..596T} and, further,
evidence for the momentum of hot upflowing plasma balancing the
momentum of cool, downflowing plasma during explosive evaporation is
indicated in some events \citep{2006A&A...455.1123T}.

The  detection of blueshifted emission components of hot lines was
initially confined to spatially-unresolved X-ray spectra
\citep{doschek80,antonucci82}, which always showed a dominant emission
component near the rest wavelength of the line with the high velocity
component present as a weaker shoulder on the short wavelength
side. The dominant, at-rest emission component can not be explained by
chromospheric evaporation models unless an ensemble of many events
occurring at different times is assumed
\citep{1983ApJ...265.1090C,1997ApJ...489..426H,1998ApJ...500..492H,2005ApJ...618L.157W}.
Spatially resolved spectra would be expected to resolve flare kernels
and thus reveal a dominant, highly-blueshifted plasma component but a
detailed study had to await the launch of
the Coronal Diagnostic Spectrometer (CDS) on board the Solar and
Heliospheric Observatory (SOHO) in 1995. CDS observed the \ion{Fe}{xix}
\lam592.2 line, formed at 9~MK, with a spatial resolution of around
6--8\arcsec. Many papers reported blueshifts of the \ion{Fe}{xix}
line, with upflow speeds of up to 230~\kms\ \citep{2003ApJ...586.1417B, 2004ApJ...613..580B,
  2005A&A...438.1099H,2006ApJ...638L.117M,2006ApJ...642L.169M,
  2006SoPh..234...95D}. The fairly low spectral resolution of CDS meant that
multi-component fitting of the \ion{Fe}{xix} line was generally
not possible, although \citet{2003ApJ...588..596T},
\citet{2006ApJ...638L.117M} and
\citet{2006A&A...455.1123T} reported observations for
which two components could be fit. For the latter two papers the high velocity
component was dominant and the blueshifted components implied upflow
speeds of 230 and 200~\kms. These results were the first instances
whereby a dominant upflow component was measured during the impulsive
phase of a flare, suggesting that the evaporation site had been
resolved. We note that the upflow velocities are somewhat smaller than the values
derived by multiple component fitting of  spatially unresolved X-ray data for
which values of around 400~\kms\ could be measured
\citep{antonucci82,1989ApJ...344..991F}. This may reflect the lower
temperature of formation of the \ion{Fe}{xix} line compared to the
X-ray lines.

The EIS instrument presents a significant improvement in both spatial
and spectral resolution over CDS, and it also has access to
hotter emission lines from \ion{Fe}{xxiii} and
\ion{Fe}{xxiv}, formed in the range 10--30~MK.
The first study of the impulsive phase of a flare observed by EIS was
performed by \citet{milligan09}, who presented Doppler measurements at
the footpoints of a C1.1 flare. Emission lines formed below 5~MK
were found to have Gaussian profiles, and Doppler shifts showed a
change from redshifts to blueshifts at around 2~MK. This was cited as
evidence of explosive evaporation, whereby cooler plasma recoils
towards the photosphere and hotter plasma rises upwards towards
the corona. Lines from the hottest species, \ion{Fe}{xxiii} and
\ion{Fe}{xxiv} (10--30~MK), showed two emission components, the dominant being
close to the rest wavelengths of the lines, the weaker at velocities
of $< -200$~\kms. This is similar to the earlier X-ray observations,
a surprise given that the footpoints are resolved by EIS.

The most relevant study for the present work is that of
\citet{2010ApJ...719..213W} who presented observations of a C9.7
confined flare. The region displayed four intense brightenings during
the flare rise phase that were interpreted as loop footpoints. Three
spectra of \ion{Fe}{xxiii} \lam263.77 (formed at 15~MK)  obtained of
one of the footpoints during the rise phase and 
separated by 160~s show (1) a small blueshift
of $-55$~\kms, (2) a two component profile with a
dominant component  at a velocity of at least $-382$~\kms, and (3) a single component
profile at a velocity of $-40$~\kms. For spectrum 2, the \ion{Fe}{xvi}
\lam262.99 line (3~MK) showed a two component profile with the weaker 
blueshifted component at $-116$~\kms. All cooler lines do not show
evidence of a blueshifted component in any of the spectra. 
The dominant high velocity \ion{Fe}{xxiii} upflow found in spectrum 2
is similar to that found from CDS data by \citet{2006ApJ...638L.117M} and
\citet{2006A&A...455.1123T} only that the magnitude of the velocity is
significantly larger.
We note that the
properties derived from spectrum 2 of \citet{2010ApJ...719..213W} are quite similar to
those found in the present work.

The impulsive phase of a smaller B2 class flare was studied by
\citet{2011A&A...526A...1D} and a key result was the finding of
blue-shifted components of \ion{Fe}{xiv, xv} and {\sc xvi} lines
(2--3~MK) with velocities of 40--60~\kms. The rest components of the
lines were stronger in intensity. The \ion{Fe}{xxiii} and
\ion{Fe}{xxiv} lines were weak in this flare, and velocity results
were not discussed.

\citet{2011A&A...532A..27G} presented observations of a C7 flare and
found a blue wing enhancement to the \ion{Fe}{xvi} \lam262.98 emission
line (formed at 2.5~MK), corresponding to upflow velocities of up to
140~\kms. As noted above, a blue-shifted component for \ion{Fe}{xvi} was found
by \citet{2010ApJ...719..213W} and \citet{2011A&A...526A...1D}, and is
also found in the present work. \citet{2011A&A...532A..27G} did not
find any high velocity components for other lines, and Doppler shifts
were generally in the $-20$ to 0~\kms\ range. 

The first EIS observation of a flare with a sit-and-stare study was
recently reported by \citet{2013ApJ...762..133B} for a C1 flare. Although the slit
position did not lie directly on the flare loop footpoint, line-of-sight velocities
measured in the leg of the loop from \ion{Fe}{xxiii} gave a peak
velocity of $-208$~\kms. Note that the emission line was completely
blueshifted for 156~s, and then displayed a two component profile with
a dominant stationary component and weak blueshifted component for a
further 56~s.

\citet{doschek12} studied  a M1.8 class flare observed by
EIS in 2012 March, and a spectrum obtained during the rise phase did
not reveal any significant Doppler flows in \ion{Fe}{xxiii} \lam263.77
(formed at 15~MK), however emission lines of \ion{Fe}{xiii} and
\ion{Fe}{xiv} (formed at 2~MK) showed extended short wavelength wings that are due to
downflowing plasma. Similar profiles are found in the present work for
\ion{Fe}{xiv}. Doppler shifts of $\approx$~$-100$~\kms\ for \ion{Fe}{xxiii} are found at a later
phase of the flare by \citet{doschek12} suggesting the presence of
hot, evaporating plasma.

The new aspect of the present work is the ability to accurately
co-align the EIS data with high spatial and temporal images from the
AIA instrument, allowing the flare evaporation site to be studied in
great detail.
Our analysis begins with a summary of the
data-sets used (Sect.~\ref{sect.dataset}) and an overview of the
active region and flare
(Sect.~\ref{sect.overview}). Sect.~\ref{sect.eis} presents the
analysis of the EIS flare 
kernel spectrum, and Sects.~\ref{sect.aia} and \ref{sect.mag} present
analysis of the imaging data-sets from SDO and \hinode. Results are
summarized in Sect.~\ref{sect.summary}.

\section{Dataset overview}\label{sect.dataset}

The data analyzed in the present work principally come from the EUV
Imaging Spectrometer \citep[EIS;][]{culhane07} on board the \hinode\
satellite, and the Atmospheric Imaging
Assembly \citep[AIA;][]{lemen12} on the Solar Dynamics Observatory
(SDO). Additional data come from the Helioseismic and Magnetic Imager
\citep[HMI;][]{scherrer12} on board SDO and the \hinode/X-Ray
Telescope \citep[XRT;][]{2007SoPh..243...63G}. Unfortunately the
Reuven Ramaty High Energy Solar Spectroscopic Imager (RHESSI) was in
spacecraft night during the rise phase of the flare and so no data
were available. 

Much of the data calibration and analysis performed for the present
work made use of IDL software routines that are part of the
\emph{Solarsoft} distribution.
AIA data downloaded from the Joint Science Operations Center (JSOC)
or the Virtual Solar Observatory (VSO) are provided in level-1
format which means that they have been flat-fielded, de-spiked and
calibrated. 
The AIA data shown in this work were ``re-spiked'' using the IDL
routine AIA\_RESPIKE as it was found that some of the flare kernels
were incorrectly flagged by the AIA 
de-spiking routine. The files were then processed with AIA\_PREP to
convert them to level-1.5 format, which places all of the different
AIA filter images onto the same plate scale. The HMI data were also
processed with AIA\_PREP to place them on the same plate scale as the
AIA images. 

The AIA detectors register events in ``data numbers'' (DN) such that
DN values between 0 and 2$^{14}-1$ (=16,383) can be measured in a
single exposure. During flares the maximum value can be reached,
leading to saturation, and the extent of saturation varies depending
on the sensitivity of the different AIA channels and the strength of
the radiation that the channels measure. Usually fixed exposure times for
the AIA channels are used, but during flares the exposure times can be
automatically reduced to help prevent saturation. For the present
observation, however, most of the AIA channels were badly affected by
saturation during the flare, with the 94~\AA\ channel the only one
completely unaffected. (Note that in flare conditions this channel is dominated by
\ion{Fe}{xviii} \lam93.93, formed at 7~MK.)

AIA obtains full-disk images at a 12~s cadence in seven different EUV
filters, and at a 24~s cadence in two UV filters. The filters are
identified by the wavelength of peak sensitivity, and we use the
shorthand A94, A131, etc., to refer to the filters with
wavelengths 94~\AA, 131~\AA, etc. The pixel size of the images
corresponds to an angular area of 0.6 $\times$ 0.6~arcsec$^{2}$ on the
Sun.

EIS spectroscopic observations are obtained by scanning a narrow slit
over an area of the Sun. At each pixel position along the slit,
spectra covering the ranges 170--212~\AA\ and 246--292~\AA\ are
obtained, however due to telemetry restrictions only narrow wavelength
windows centered on specific emission lines are usually downloaded.
On 2011 February 16 
EIS continuously ran a single observation study
called HH\_Flare\_180x160\_v2 for the period 06:17--12:23~UT, yielding
64 rasters in all. For each raster the 2\arcsec\ slit scanned an area of 180 $\times$
160~arcsec$^2$  in 5~min 45~s. An 8~s exposure time was used and the slit
jumped 5\arcsec\ between exposure positions. Ten wavelength windows are
obtained with the HH\_Flare\_180x160\_v2 study, and the particular  emission
lines studied  in the present work are listed in Table~\ref{tbl.eis}.

The EIS data were calibrated with the EIS\_PREP routine in \emph{Solarsoft}
using the standard processing options listed on the EIS
wiki\footnote{http://msslxr.mssl.ucl.ac.uk:8080/eiswiki/}, and ``missing''
data (due to warm pixels, cosmic rays, etc.) were interpolated using
the procedure described in \citet{eis_sw13}. The /CORRECT\_SENSITIVITY
keyword was used, which implements a wavelength-independent sensitivity
decay for the instrument with an e-folding time of 1894~days, which
 means that the  line intensities tabulated in the present work are
a factor 2.3 higher than if the pre-launch calibration had been
used. At the time of writing the EIS calibration is under revision by
the EIS team, and it seems likely that some wavelength regions have
decayed less strongly than suggested by the earlier analysis. If
correct, then this
would lower the intensities given in the present work.

Given the high time cadence of the instruments used for this work, it
is important to state observation times precisely. When times of
individual exposures of AIA, HMI and EIS are referred 
to, the time corresponds to the midpoint of the exposure given in
Coordinated Universal Time (UTC). For the SDO
instruments, this is the parameter T\_OBS stored in the data
header, although for HMI this value has to be converted from
International Atomic Time to UTC.
For EIS the midpoint of an
exposure is defined as the time midway between the shutter open and
close times, which are stored in the EIS data file headers.

\section{Overview of active region of AR 11458}\label{sect.overview}

Active region AR 11458 was the first major flaring active region of
solar cycle 24 and so many aspects of the region's flares and
magnetic evolution have already been studied. The dominant flaring activity
took place during 2011 February 13--18, with the largest event an X2.2
class flare on February 15 that produced a sunquake
\citep{2011ApJ...734L..15K} and Earth-directed coronal mass ejection
\citep[CME;][]{2011ApJ...738..167S}. A M1.6 flare on February 16 also
produced a CME that has been spectroscopically studied with the EIS
instrument \citep{harra11, veronig11}. 

Prior to 15~UT on February 15 the flares from AR 11458 were generally
eruptive events with significant jetting and CME activity. From 15~UT
up to the M1.6 flare at 14:25~UT on February 16, however, the five C5
or greater flares that took place were all confined flares.
For this work we focus on the fourth flare in this sequence: a M1.1
class flare that peaked in the GOES 
1--8~\AA\ X-ray light curve at 07:44~UT on  February 16. This
flare was chosen as the EIS observation revealed the location of a high
velocity upflow at temperatures of 10--30~MK during the flare rise
phase. Such locations are often missed by EIS raster scans as they
have short lifetimes and they occur in only specific parts of
the active region. The topology of the active region and post-flare
loops also gives a relatively clean view of the upflow site, with a
clear correlation with intense, compact brightenings in AIA
images. The EIS spectroscopic properties of the upflow site are
similar to those measured by \citet{2010ApJ...719..213W} from a C9.7
flare observed in 2007, but the availability of SDO data gives a
greater insight into the properties of the flare upflow region.

\begin{figure}[h]
\epsscale{0.9}
\plotone{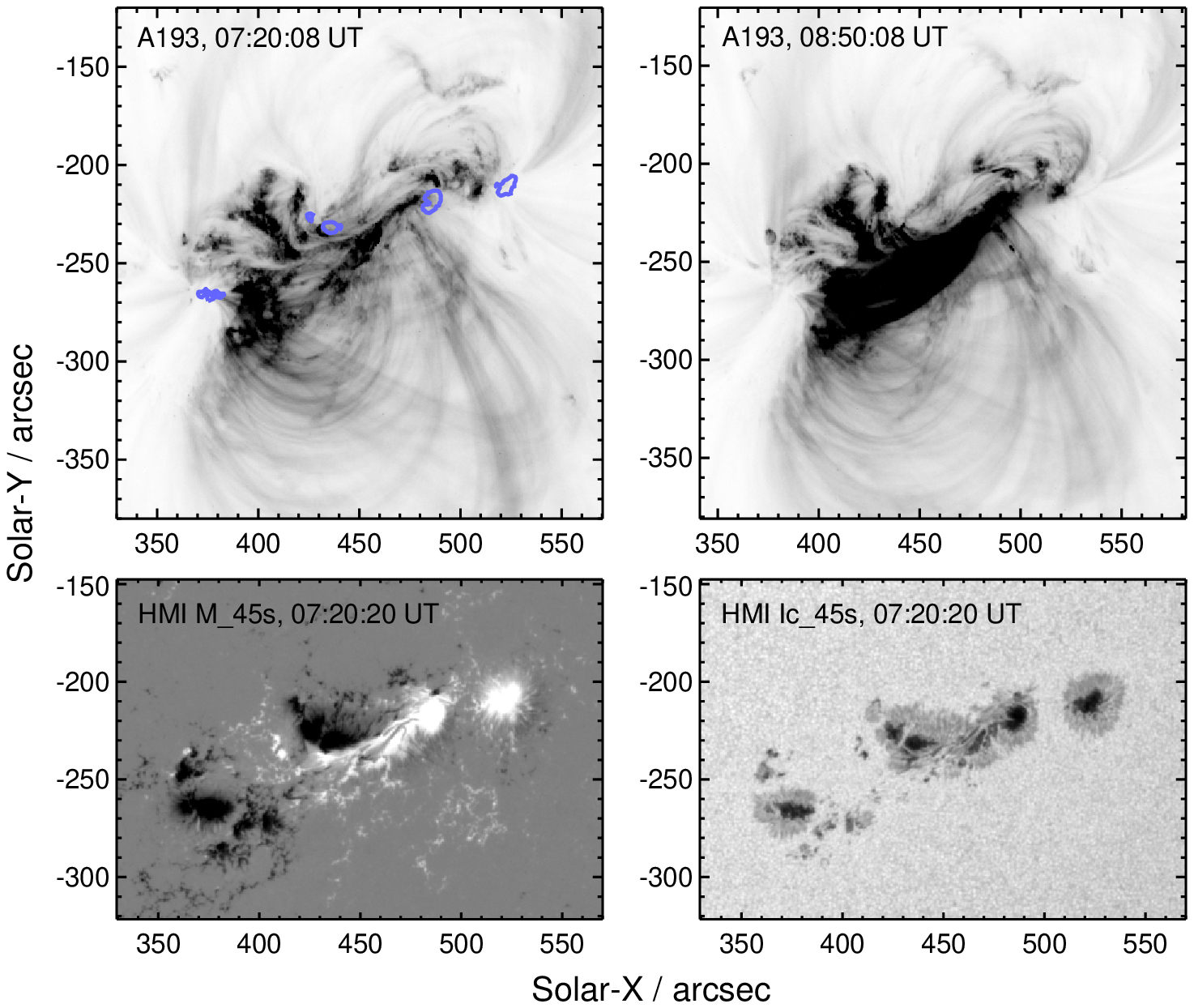}
\caption{The upper panels show AIA 193~\AA\ images obtained before and
after the 2011 February 16 07:44~UT flare. A reverse intensity scaling has been applied
and the maximum intensity value is 2500~DN~s$^{-1}$ in both
images. The lower panels show HMI LOS magnetogram (left) and white
light continuum (right) images from before the flare. The blue
contours on the top-left image show the locations of the sunspot umbrae.}
\label{fig.ar}
\end{figure}

Figure~\ref{fig.ar}
shows before-and-after images of AR 11458 obtained with the AIA 193~\AA\ filter,
which demonstrate that the peripheral loop structures remain largely
unchanged following the flare. The HMI LOS magnetogram and continuum
images in Figure~\ref{fig.ar} show the photospheric structure of the
active region prior to the flare. 

\begin{figure}[h]
\epsscale{1.0}
\plotone{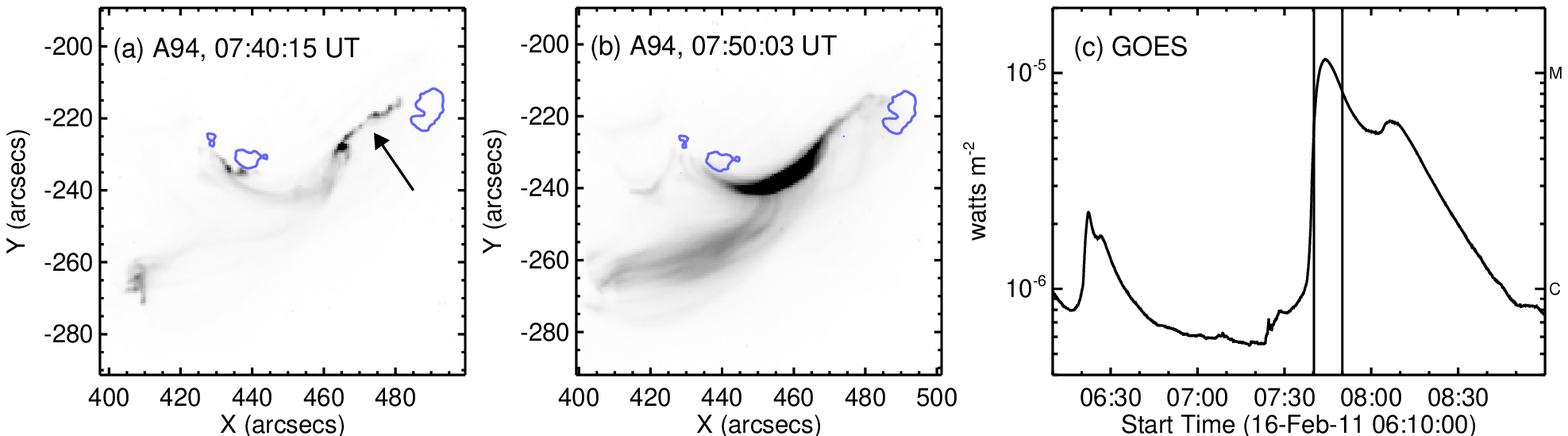}
\caption{Panels a and b show AR 11458 at two different times, as seen in
the AIA 94~\AA\ filter. A negative intensity scaling is used, with
black corresponding to the brightest regions. Both images have been
saturated to a level of 3000~DN~s$^{-1}$; the actual intensity maxima
are 5301 and 5328~DN~s$^{-1}$. An arrow indicates a group of flare kernels, one of which is studied in the present work. Blue
contours show the location of sunspot umbrae as determined from
co-temporal HMI data. Panel c
shows the GOES 1--8~\AA\ light curve, with two vertical lines 
representing the times at which the A94 images were obtained.}
\label{fig.goes}
\end{figure}

The flare emission is best seen in the AIA 94~\AA\ filter images and
Figure~\ref{fig.goes} shows two images, one from the rise phase (a) and one
taken a few minutes after the flare peak (b). The image times are
indicated on the GOES X-ray light curve (Figure~\ref{fig.goes}c). During
the rise phase,  flare kernels are seen at three
locations in the active region that approximately correspond to the
footpoints of the post-flare loops that eventually appear. The most
intense flare loops connect the two central sunspots (see also
Figure~\ref{fig.ar}), while another fainter, more twisted set of loops
connects the central, positive polarity sunspot to the negative
polarity sunspot at the east side of the active region
(Figure~\ref{fig.goes}b).

\section{EIS data analysis}\label{sect.eis}

\begin{figure}[h]
\epsscale{1.0}
\plotone{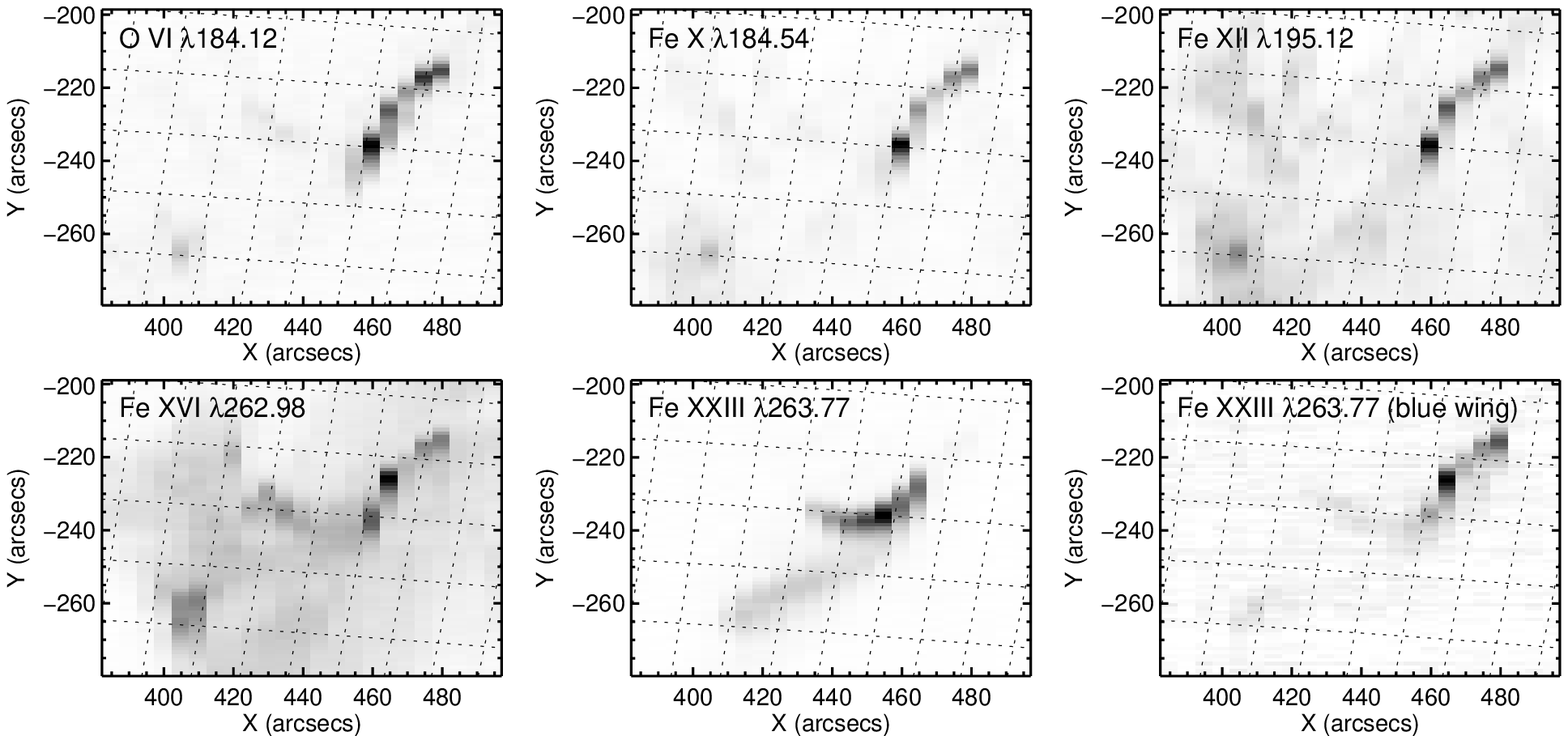}
\caption{Six images from the EIS raster that began at 07:38:36~UT. A
  reverse intensity scaling has been applied to each image. The grid
  on each image shows lines of latitude and longitude, spaced at $1^{\circ}$ intervals. The leftmost line of longitude is $+24^{\circ}$,
  and the lowermost line of latitude is $-22^{\circ}$. The
  \ion{Fe}{xxiii} \lam263.77 blue wing corresponds to \ion{Fe}{xxiii}
  upflow velocities of around 350~\kms. See main text for more details.}
\label{fig.eis-ims}
\end{figure}

The group of A94 flare kernels highlighted in Figure~\ref{fig.goes}a
were scanned by EIS with a raster that began at 07:38:26~UT and finished
at 07:44:22~UT (note that EIS
rasters west-to-east). Five consecutive raster positions obtained
between 07:40:16~UT and 07:40:51~UT revealed five intense, compact
brightenings in most of the EIS lines (Figure~\ref{fig.eis-ims}).
The hot \ion{Fe}{xxiii} \lam263.77 line (formed at 15~MK) does not
show the brightenings as a short, intense post flare loop dominates
the image. However an image formed in the short wavelength wing of the
line at around $-350$~\kms\ does show the brightenings (lower-right
panel of Figure~\ref{fig.eis-ims}).
For
this work we focus on the right-most of the five brightenings, which
was observed by EIS at 07:40:16~UT.

\begin{figure}[h]
\epsscale{1.0}
\plotone{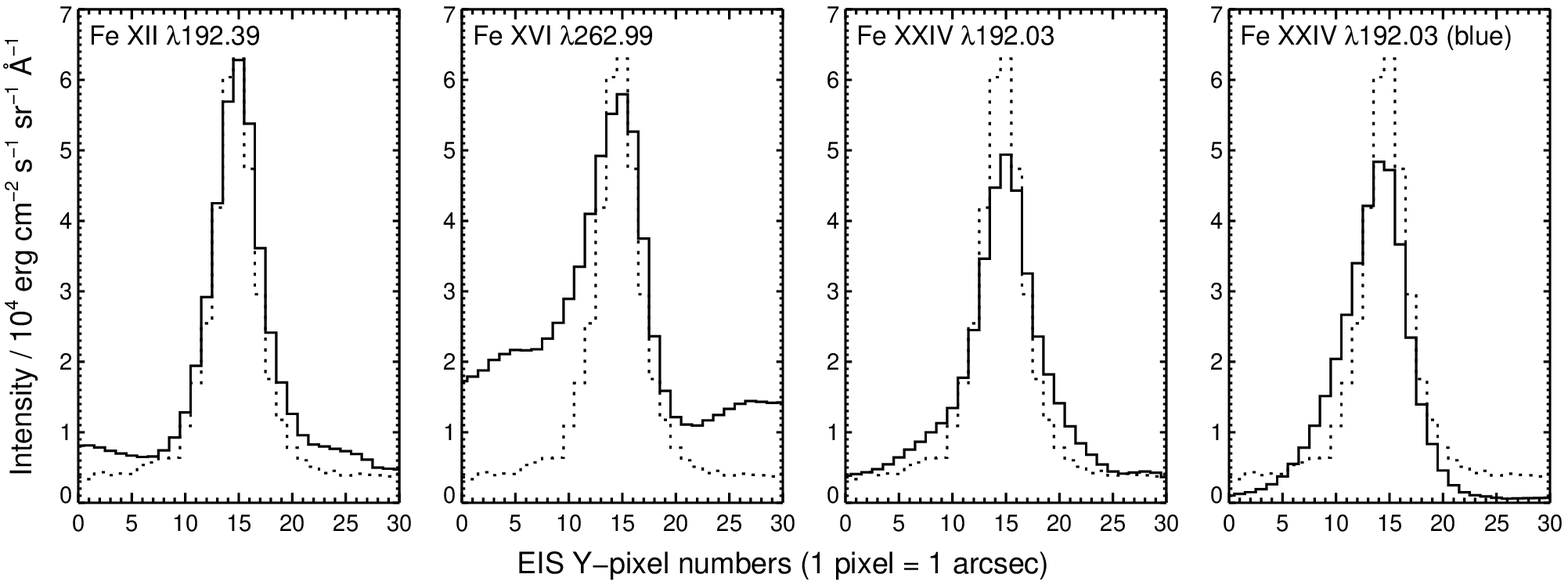}
\caption{The dotted lines in each plot show the intensity cross-section
of \ion{O}{vi} \lam184.11. The solid lines show the intensity
cross-sections of the four species identified in the top-left corner
of each plot. The
\ion{Fe}{xxiv} \lam192.03 (blue) intensity was obtained at a
wavelength of 191.77~\AA, corresponding to a \lam192.03 velocity of
$-400$~\kms. The intensity of the \lam192.03 (blue) cross-section has
been divided by a factor 2. No scaling has been applied to the other
cross-sections.}
\label{fig.eis-xs}
\end{figure}

Figure~\ref{fig.eis-xs} shows intensity cross-sections through the selected
brightening in the solar-Y direction (i.e., along the EIS slit). Each
intensity cross-section is co-spatial in the X-direction and
co-temporal by the nature of spectrometer observations. The
cross-sections have been aligned with each other by making use of the
known spatial offsets in the solar-Y direction (due to grating tilt
and CCD spatial offsets) that are obtained from the IDL routine
EIS\_CCD\_OFFSET. 

A striking feature of the brightening is that the intensity peaks at
the same pixel for all temperatures from 0.3~MK (\ion{O}{vi}) to 20~MK
(\ion{Fe}{xxiv}). The intensity profiles are also very similar: 
Gaussian shapes with a full-width at half-maximum of about 4
pixels. (The \ion{Fe}{xvi} profile is affected by significant
background coronal emission.) The EIS spatial resolution has been
independently estimated as 3--4\arcsec\ based on comparisons with AIA
images and studies of transition region brightenings\footnote{See
  discussion at \url{http://msslxr.mssl.ucl.ac.uk:8080/eiswiki}.},
suggesting the observed feature is not resolved by EIS. This is
confirmed by the higher spatial resolution images from AIA (Sect.~\ref{sect.aia}).

At the location of the brightening it is found that \ion{Fe}{xxiv}
\lam192.03 has a strong emission component at a velocity of $\approx$
$-400$~\kms\ -- see discussion later. The intensity cross-section for
this blue component is also shown in Figure~\ref{fig.eis-xs} and is
consistent with the others.

The presence of plasma with a wide range of temperatures  compacted into a
single bright feature is expected for the standard model of
chromospheric evaporation whereby chromospheric plasma at the
footpoint of the flare loop is rapidly heated to multi-million degree
temperatures. Further support comes from the presence of rapidly
upflowing plasma at the hottest temperatures, however the finding of
multiple velocity components in the emission lines is not compatible
with a simple, single loop model and suggests further complexity.

In the following sections we study the spectroscopic properties of the
brightening in more detail. To do this we create a single spectrum for
the brightening in order to measure line intensities, widths and
Doppler shifts. The process is complicated however by the fact that
the background coronal emission is significant for some of the
emission lines. Sect~\ref{sect.extract} describes how the background
emission is dealt with and how the spectrum is
extracted. Sect.~\ref{sect.fitting} discusses blending issues for the
emission lines, and Sects.~\ref{sect.vel}--\ref{sect.dens}
present the results for Doppler shifts, line widths, emission measure
and density.

\begin{deluxetable}{cccccccc}
\tablecaption{EIS emission line parameters.\label{tbl.eis}}
\tabletypesize{\footnotesize}
\tablehead{
  &
  &
  &
  &
  &
  &
  \colhead{Non-thermal} &
  \colhead{Log$_{10}$ (Column} \\
  \colhead{Ion} &
  \colhead{Wavelength} &
  \colhead{Log\,$T_{\rm mem}$} &
  \colhead{Component} &
  \colhead{Intensity} &
  \colhead{LOS velocity} &
  \colhead{velocity} &
  \colhead{emission measure} \\
  &
  &
  &
  &
  \colhead{(\ecss)} &
  \colhead{(\kms)} &
  \colhead{(\kms)} &
  \colhead{/cm$^{-5}$)} \\
 
}
\startdata
\ion{O}{vi} & \lam184.12 &  5.51 & 1
  &$   43751\pm   921$ & $  34.5\pm    3.8$ & $  60.0\pm    3.5$ &  29.91 \\
\noalign{\smallskip}
\ion{Fe}{x} & \lam184.54 &  6.05 & 1
  &$   82385\pm  1382$ & $  17.4\pm    3.8$ & $  81.5\pm    3.4$ &  29.65 \\
\noalign{\smallskip}
\ion{Fe}{xii} & \lam192.39 &  6.20 & 1
  &$   24687\pm   565$ & $ -19.5\pm    2.0$ & $  18.6\pm    3.2$ &  29.46 \\
     &&&2
  &$    9445\pm   697$ & $  73.7\pm    3.7$ & $  18.6\pm    3.2$ &  29.04 \\
\noalign{\smallskip}
\ion{Fe}{xiv} & \lam264.79 &  6.30 & 1
  &$   82449\pm   882$ & $ -31.0\pm    1.2$ & $  28.1\pm    2.1$ &  29.61 \\
     &&&2
  &$   21558\pm   878$ & $  64.9\pm    2.8$ & $  28.1\pm    2.1$ &  29.03 \\
\noalign{\smallskip}
\ion{Fe}{xiv} & \lam274.20 &  6.30 & 1
  &$   33731\pm   599$ & $ -33.0\pm    1.3$ & $  32.1\pm    2.2$ &  29.60 \\
     &&&2
  &$   10653\pm   567$ & $  69.0\pm    3.0$ & $  32.1\pm    2.2$ &  29.10 \\
\noalign{\smallskip}
\ion{Fe}{xv} & \lam284.16 &  6.34 & 1
  &$   18971\pm  1530$ & $-149.8\pm    3.7$ & $  29.8\pm    2.0$ &  28.29 \\
     &&&2
  &$  320560\pm  3060$ & $ -48.3\pm    0.6$ & $  29.7\pm    2.0$ &  29.52 \\
     &&&3
  &$   40236\pm  1310$ & $  79.9\pm    1.6$ & $  29.7\pm    2.0$ &  28.62 \\
\noalign{\smallskip}
\ion{Fe}{xvi} & \lam262.98 &  6.43 & 1
  &$     972\pm   297$ & $-249.0\pm   18.4$ & $  20.4\pm    3.0$ &  28.22 \\
     &&&2
  &$    6398\pm  1045$ & $-145.6\pm    8.2$ & $  20.4\pm    3.0$ &  29.04 \\
     &&&3
  &$   31990\pm  1188$ & $ -59.9\pm    2.2$ & $  20.3\pm    3.0$ &  29.74 \\
     &&&4
  &$    1381\pm   334$ & $  79.0\pm   12.4$ & $  20.3\pm    3.0$ &  28.38 \\
\noalign{\smallskip}
\ion{Fe}{xxiii} & \lam263.77 &  7.16 & 1
  &$   10584\pm   470$ & $-384.7\pm    3.8$ & $ 120.6\pm    6.6$ &  30.15 \\
     &&&2
  &$    3985\pm   382$ & $  24.0\pm    8.1$ & $ 111.6\pm   13.3$ &  29.73 \\
\noalign{\smallskip}
\ion{Fe}{xxiv} & \lam192.03 &  7.25 & 1
  &$  102900\pm   685$ & $-404.8\pm    1.3$ & $ 113.3\pm    3.0$ &  29.89 \\
     &&&2
  &$   42488\pm   599$ & $   5.3\pm    1.7$ & $ 127.1\pm    3.6$ &  29.51 \\
\noalign{\smallskip}
\ion{Fe}{xxiv} & \lam255.11 &  7.25 & 1
  &$   26762\pm  1172$ & $-425.1\pm    2.6$ & $  85.3\pm    5.7$ &  29.70 \\
     &&&2
  &$    6856\pm   738$ & $  -4.6\pm    4.6$ & $  34.8\pm   10.3$ &  29.11 \\

\enddata
\end{deluxetable}

\subsection{Extracting the spectrum of the flare kernel}\label{sect.extract}

As discussed in the previous section, EIS does not spatially resolve
the flare kernel site and so we assume that the Gaussian-shaped
intensity profiles shown in Figure~\ref{fig.eis-xs} all come from a
single, unresolved structure. Practically this means that we sum the
intensity from 9 Y-pixels distributed over the intensity cross-sections
to yield a single spectrum for the flare kernel. However,
Figure~\ref{fig.eis-xs} shows that there is background emission 
against which the kernel appears in all of the EIS lines. This
background emission is particularly significant for \ion{Fe}{xv} and
\ion{Fe}{xvi}. In addition to affecting the intensity of the emission
lines, we also find that this emission distorts the emission line
profiles, which then affects the measurement of line widths and
velocities. For this reason we subtract a pre-flare spectrum from the
flare kernel spectrum.

The pre-flare spectrum was obtained from the previous EIS raster,
which began at 07:32~UT. This raster
showed much
reduced emission at the positions of the five flare kernels and so we
can consider it to give a good representation of the pre-flare corona
at the locations of the brightenings.

The background corona subtraction is performed with the
\emph{Solarsoft} routine EIS\_MASK\_SPECTRUM,
which performs the procedure as follows. Each spectral window from an
EIS raster yields a 3D intensity array of wavelength, solar-X and solar-Y
pixels. Since the satellite pointing does not change between the pre-flare and
flare rasters, then potentially one can simply take the spatial pixels
corresponding to the brightening and subtract the pre-flare spectra
from the flare spectra. The situation is complicated, however, by the
fact that the EIS  wavelength scale drifts with time due to the
thermally-induced motion of the EIS grating during an orbit (referred
to as spectrum drift). Therefore in order to perform the pre-flare
spectrum subtraction, it is necessary to place the spectra at each
spatial pixel in each raster onto a common wavelength scale. This is
done with the IDL routine EIS\_SHIFT\_SPEC, which shifts the spectra for each
spatial pixel of  a
raster onto a common wavelength scale by making use of the
spectrum drift and slit tilt corrections of
\citet{2010SoPh..266..209K}. The pre-flare spectrum can then be
subtracted from the flare spectrum at each spatial pixel. The
background-subtracted spectra from the 9 Y-pixels that span the flare
kernel are then summed  to
yield the final spectrum.

\begin{figure}[h]
\epsscale{1.0}
\plotone{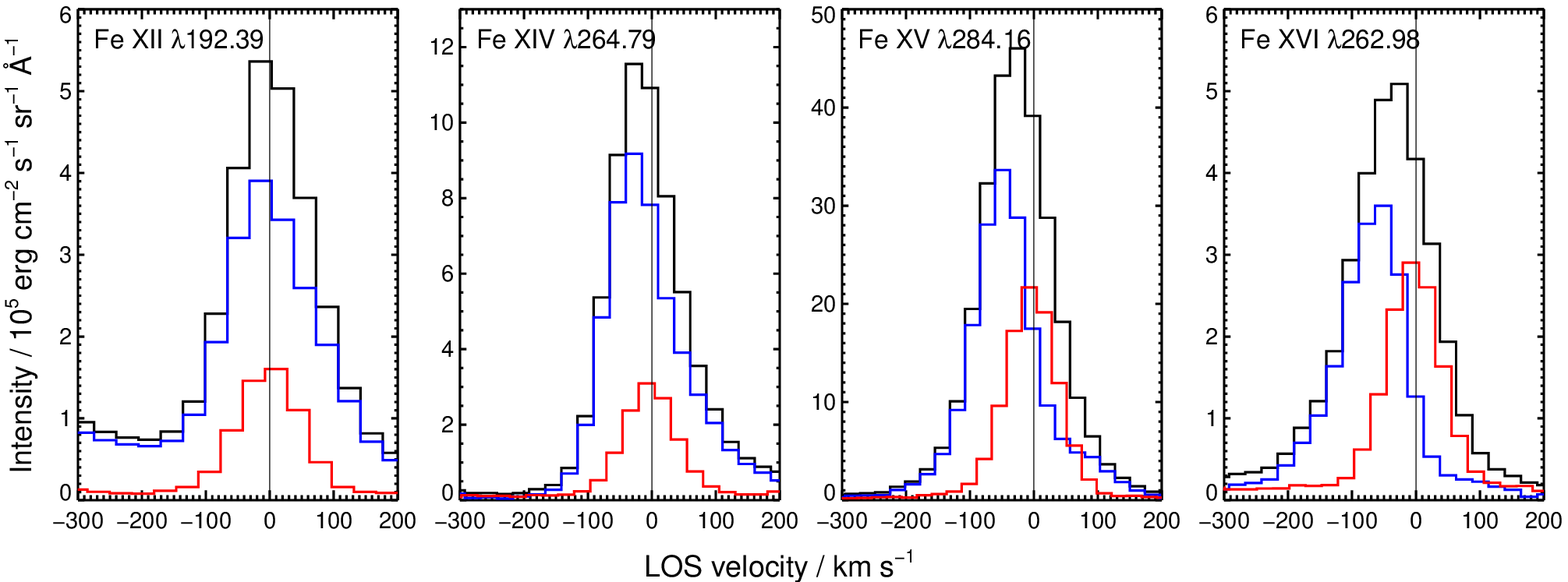}
\caption{Four emission lines observed by EIS are shown. The red
  lines show the pre-flare spectrum, the black lines the flare kernel
  spectrum, and the blue lines the spectrum obtained by subtracting
  the pre-flare spectrum from the flare spectrum.}
\label{fig.sub-spec}
\end{figure}

Figure~\ref{fig.sub-spec} shows how the subtracted spectrum (blue)
compares with the un-subtracted spectrum (black) and the pre-flare
spectrum (red) for four emission lines. For \ion{Fe}{xv} \lam284.16
and \ion{Fe}{xvi} \lam262.98 it is clear that the pre-flare emission lines,
which are close to at rest, when subtracted result in the centroid
being pushed to shorter wavelengths. The method for setting the rest
wavelength for these spectra is discussed in Sect.~\ref{sect.vel}.

A key uncertainty in the background subtraction method is the EIS
pointing stability: if the pointing changed between the 07:32 and
07:38~UT rasters, then the pre-flare background at the site of the
brightening is no longer valid.  The \hinode\ pointing is known to
exhibit peak-to-peak pointing fluctuations of up to 3\arcsec\ in both
X and Y directions \citep{2010ApJ...713..573M}, but a benefit for the
present analysis is that \hinode\ was pointed at AR 
11458 for long periods of time, meaning that the satellite and the
instruments received a nearly fixed illumination from the Sun
(re-pointing of the satellite leads to varying illumination and thus
thermal effects). The IDL routine EIS\_JITTER returns estimates of the
instrument pointing jitter, and this shows variations of $\le
0.25$\arcsec\ in solar-X and $\le 1.2$\arcsec\ in solar-Y. For
the solar-Y direction, pointing can be checked by
comparing intensity cross-sections and it can be seen that features in
solar-Y away from the flare site are well-matched to within a pixel
between the 07:32 and 07:38~UT rasters. We thus believe our background
subtraction method is not significantly affected by pointing jitter.

\subsection{Line fitting notes}\label{sect.fitting}

The spectrum extraction method described above yields a single EIS
spectrum for the flare kernel. The emission lines were fit with
Gaussian functions using the IDL routine SPEC\_GAUSS\_EIS.
Most of the lines have non-Gaussian shapes, suggesting the
presence of multiple plasma components at different velocities and
specific details on how each emission line was treated are given
below. The line fit parameters, expressed as intensity, line-of-sight
(LOS) velocity and non-thermal broadening, are given in
Table~\ref{tbl.eis}. From the line intensity the column emission
measure can be derived, and this quantity is also shown in
Table~\ref{tbl.eis}. The temperature of maximum emission, $T_{\rm
  mem}$, given in Table~\ref{tbl.eis} is the temperature at which a line's
contribution function peaks and is derived using version 7.1 of the
CHIANTI database \citep{dere97,chianti71}.

The error estimates shown in Table~\ref{tbl.eis} ultimately derive from
photon statistics errors, which are generally small due to the
strength of the emission lines. As discussed earlier there is some
uncertainty in the absolute calibration of EIS and this is not
reflected in  the intensity uncertainties. The lack of an absolute
wavelength calibration for EIS means that there is a systematic
uncertainty of $\approx$~10~\kms\ (see Sect.~\ref{sect.vel}) that is
not included in Table~\ref{tbl.eis}. The uncertainty in the
non-thermal broadening is discussed in Sect.~\ref{sect.wid}. The
column emission measure is directly proportional to the line
intensity, and so the uncertainty (not displayed in
Table~\ref{tbl.eis}) can be obtained from the line
intensity uncertainty.

The two coolest lines, \ion{O}{vi} \lam184.12 and \ion{Fe}{x}
\lam184.54, were both fit with single Gaussians. The former is
well-isolated in the spectrum, but the line's doublet partner at
183.94~\AA\ is present at the edge of the window and masks out any
high velocity (200--300~\kms) component that may be present. However
the partial line that is present is consistent with the expected
\lam183.94/\lam184.11 intensity ratio, suggesting that there is no
significant high velocity component.

\ion{Fe}{x} \lam184.54 is blended on the short wavelength side with
\ion{Fe}{xi} \lam184.41, which is noticeably stronger (relative to \lam184.54)
than in typical active region conditions, likely due to the high
density of the brightening. The two
lines were fit simultaneously with two Gaussians, and fits were found
to be good, with in particular no suggestion that \lam184.54 is asymmetric.

The emission lines of \ion{Fe}{xii--xvi} all show non-Gaussian
profiles and they have been fit with multiple Gaussians. As EIS
emission lines typically have a full-width at half-maximum of about 3
pixels then they are not ideally suited for detailed modeling in this
manner. In particular, the parameters of weak components can be quite
uncertain. For this work, we force the multiple Gaussian components
for a single line to have the same width. While there is no physical
justification for this, it does provide a baseline against which the
plasma components of different ions can be compared and the results
described later suggest that this is valid.

\ion{Fe}{xii} \lam192.39 line is very close to the edge of its
wavelength window and also lies on a sloping background
level resulting from the nearby strong \ion{Fe}{xxiv} \lam192.03
line. The line shows a clear asymmetry, with a steeper blue side to the
profile, and has been fit with two Gaussians. The stronger \lam195.12
line also shows a clear asymmetry, but the profile is complicated by
the presence of \ion{Fe}{xii} \lam195.18 in the red wing. This line
becomes quite strong at high densities \citep[see, e.g.,][]{young09}
and thus, if the two Gaussian model for \lam192.39 is assumed for
\lam195.12 then \lam195.18 would enhance the component on the red
side of  the line. Fitting two Gaussians to the profile confirms that
this is the case and, given the difficulties in deconvolving the blend
we choose not to present the parameters for \lam195.12.

The density sensitive \ion{Fe}{xiv} lines  at 264.79 and
274.20~\AA\ are both blended: \lam264.79 with \ion{Fe}{xi} \lam264.77,
and \lam274.20 with \ion{Si}{vii} \lam274.18. It is not possible to
directly estimate the contributions of either of these lines to the
\ion{Fe}{xiv} lines from the flare spectra, however the authors have
studied complete EIS spectra of an active region and find that the
contributions are typically $<5$\%\ when the \ion{Fe}{xiv} lines are
strong. We therefore believe it is reasonable to neglect these
blending species.
Both \ion{Fe}{xiv} lines show clear asymmetries 
such 
that the long wavelength sides of the profiles are less steep
than the short wavelength sides. The \lam274.20 long wavelength wing
extends beyond the edge of the wavelength window (which was only 16
pixels wide) and a two Gaussian
fit with a constant background was used. We note that the \lam274.20
profile is very similar to that found by \citet{doschek12} during the rise
phase of a M1.8 flare. The \lam264.79 wavelength
window is wider than that for \lam274.20 but the long wavelength wing is still found to extend
beyond the window edge, and a three Gaussian fit was performed with a
constant background. The third, weakest Gaussian is at a velocity of
$+166$~\kms. There is a nearby \ion{Fe}{xvi} line at 265.00~\AA\
which, if it is assumed to have the same velocity as the dominant
\ion{Fe}{xvi} \lam262.98 component, would place it at $+180$~\kms\ in
the \ion{Fe}{xiv} \lam264.79 reference frame. The \ion{Fe}{xvi}
\lam265.00/\lam262.98 ratio has a fixed value of 0.096 based on atomic
physics parameters, while the intensity ratio of the third Gaussian to
the dominant intensity component of \lam262.98 is 0.16. We thus
conclude that the third Gaussian component to \ion{Fe}{xiv} \lam264.79
is dominated by \ion{Fe}{xvi} and so we do not list it in
Table~\ref{tbl.eis}. The two Gaussian fits to \lam264.79 and
\lam274.20 lead to very similar LOS velocities and non-thermal
velocities for the two components, giving confidence that the fits
accurately model the line profiles. The intensity ratios of the two
components are different, however, which reflects different densities
for the two plasma components -- this is discussed in Sect.~\ref{sect.dens}.

\ion{Fe}{xv} \lam284.16 shows a dominant intensity component with two
weaker components either side and was fit with three Gaussians with
the same width.
\lam284.16 is known to be partly blended with
\ion{Al}{ix} \lam284.03, but this is negligible for the present
spectrum (\lam284.16 is the strongest line by intensity of the lines measured).

\ion{Fe}{xvi} \lam262.98 shows a similar line profile to \ion{Fe}{xv}
\lam284.16 with a dominant intensity component and weaker emission
on each side. The short wavelength emission could not be accurately fit
with a single Gaussian and so two Gaussians were necessary, giving
four in all. A
weak, unknown line is found in EIS spectra at 262.70~\AA\
\citep{brown08}, corresponding to a velocity of $-320$~\kms\ which is
too far away to account for the \ion{Fe}{xvi} component at
$-249$~\kms. We note that \citet{2011A&A...526A...1D} showed a
\ion{Fe}{xvi} \lam262.98 line profile for a flare kernel that also
displayed an extended blue wing that was fit with two Gaussian
components corresponding to upflow speeds of 60 and 170~\kms.

\begin{figure}[h]
\epsscale{1.0}
\plotone{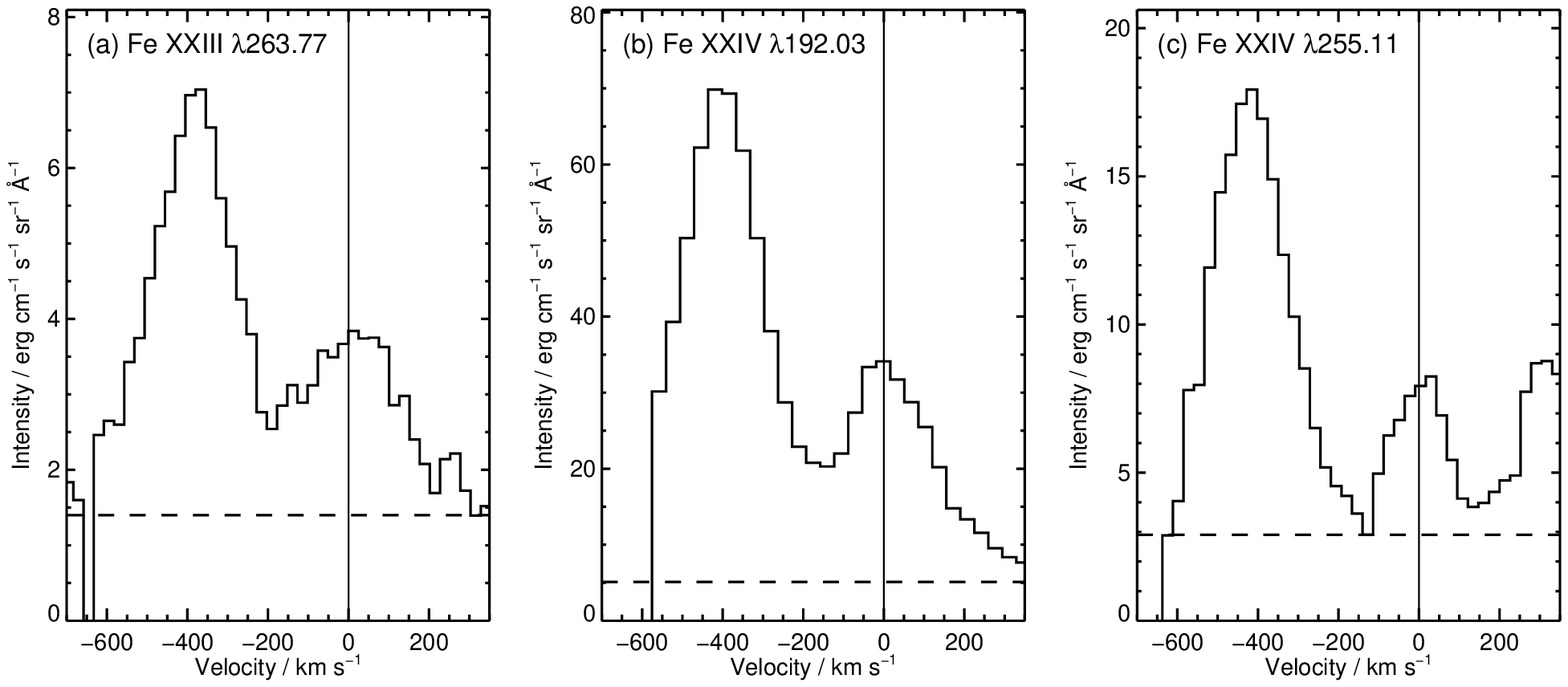}
\caption{Emission line profiles from the flare kernel spectrum for (a)
\ion{Fe}{xxiii} \lam263.77, (b) \ion{Fe}{xxiv} \lam192.03, and (c)
\ion{Fe}{xxiv} \lam255.11. The dashed lines indicate the background
level in the spectra.}
\label{fig.hot-profiles}
\end{figure}

The \ion{Fe}{xxiii} and \ion{Fe}{xxiv} line profiles are the most
striking in the flare kernel spectrum as they show strong blue-shifted
emission at around $-400$~\kms\ in addition to a weaker component near
the rest velocities of the lines (Figure~\ref{fig.hot-profiles}). We
note that the \ion{Fe}{xxiii} \lam263.77 profile is quite similar to
that shown by \citet{2010ApJ...719..213W} from a C9.7 flare
kernel. The approximate background levels in the 
spectrum, as determined from the minimum intensity within each line's
wavelength window, are also shown in Figure~\ref{fig.hot-profiles}. The
narrowness of the wavelength windows used for 
the study means that the background levels are likely to be
over-estimates of the real background. 

The similarity of the line profiles for all three lines provides
strong evidence that the highly-blueshifted components are real and
not due to strong emission in emission lines on the short wavelength
sides of each line, but we briefly discuss known blending issues for
each.

\citet{2011A&A...526A...1D} demonstrated that two weak lines blend
with \ion{Fe}{xxiii} \lam263.77 between $-50$ and $-150$~\kms\
(relative to \lam263.77). Figure~13 of their work also showed that
there is another line at about $-350$~\kms\ that potentially could
compromise the high velocity component found here. This line is found in quiet Sun
and active region spectra \citep{brown08}, so it is probably formed
at temperatures of $\log\,T=6.0$--6.2. From an active region spectral
atlas 
obtained on 2011 February 20, we find that \ion{Fe}{xii} \lam192.39 is
around a factor 170  times more intense than the line at 263.4~\AA. From
Table~\ref{tbl.eis}, \lam192.39 is only a factor 3
stronger than the line at 263.4~\AA\ (interpreted
as the blue wing of \lam263.77) giving strong evidence that the strong
blue-shifted component seen in Figure~\ref{fig.hot-profiles}(a) is in fact
due to \ion{Fe}{xxiii}.

Several lines blend directly with \ion{Fe}{xxiv} \lam192.03, but when
the flare line becomes strong it easily overwhelms these lines. The
blending lines are \ion{Fe}{vii} \lam192.01, \ion{Fe}{viii}
\lam192.04, \ion{Fe}{xi} \lam192.02 and an unknown line at 192.09~\AA\
\citep{2009ApJ...707..173Y,2009A&A...508..513D,2010A&A...514A..41D}, and a useful guide to
the extent of the blending can be made by considering the strength of
the nearby \ion{Fe}{xii} \lam192.39. The combined blending lines
between 192.01 and 192.09~\AA\ can exceed the strength of the
\lam192.39 line in some circumstances \citep{landi09}, but generally
they are much weaker. It is not possible to estimate the extent of
blending in the present spectra, but it is possible that they make a
significant contribution to the rest component of the
\ion{Fe}{xxiv}. The width of the line is consistent with formation
from a very hot ion, however.  On the
short wavelength side of the profile there is \ion{Fe}{xiv} \lam191.81
which blends with the highly-blueshifted component of the
\ion{Fe}{xxiv} line. The CHIANTI \ion{Fe}{xiv} model predicts that
this line is about 2\%\ of the strength of \ion{Fe}{xiv}
\lam274.20 and so based on the intensities in Table~\ref{tbl.eis} this
line can be shown to be negligible.

\ion{Fe}{xxiv} \lam255.11 has several lines nearby, and the
narrow width of the EIS observing window makes it difficult to fit the
spectrum. The lines near to \lam255.11 are an unknown line at
254.70~\AA, \ion{Fe}{xvii} \lam254.86,
\ion{Fe}{viii} \lam255.10,
\ion{Fe}{viii} \lam255.37 and \ion{Fe}{x} \lam255.39. The
\ion{Fe}{xvii} line is particularly important when studying
blue-shifted components of the \ion{Fe}{xxiv} line as it becomes
strong in flares and lies within the blue wing, however there is no
evidence that the line is significant in the present spectrum, and so
the observed spectral features are fit with two Gaussians to represent
the rest and highly-blueshifted components of \ion{Fe}{xxiv} \lam255.11.
The similarity of the measured velocities to those of \ion{Fe}{xxiv}
\lam192.03 give confidence in the fit, although the non-thermal
velocities are somewhat lower. This could be due to the fact that the
background level is difficult to estimate due to the many blending
lines.

We highlight here the fact that the \ion{Fe}{xxiv}
\lam192.03/\lam255.11 ratios for the two velocity components
 vary significantly from the value 
expected from atomic theory. The CHIANTI atomic model gives an
expected ratio of 2.5, which is independent of density and only very
slightly temperature dependent. The ratios found here are 3.9 and 6.2
for the blue-shifted and rest components, respectively. The rest
component of \lam192.03 could be over-estimated because of the
blending noted earlier, while the blending lines for \lam255.11 could
also lead to over-estimation of this line's components although we
feel this is likely to be offset by an over-estimate of the spectrum
background level.

To investigate the \ion{Fe}{xxiv} ratio further, flare data
from 2011 February 16 obtained after the flare peak were studied. At such times the
\ion{Fe}{xxiv} emission shows  little dynamics (in
terms of broadening or Doppler shifts)  and is very strong  so it should be free from blending. We found \lam192.03/\lam255.11 ratios ranging from
4.7 to 5.1 in six spectra, again significantly above the theoretical
ratio. G.~Del Zanna has suggested in several papers
\citep{2009A&A...508.1517D,2011A&A...533A..12D,2012A&A...537A..38D}
that there is a calibration problem for EIS whereby lines near
250~\AA\ are too weak relative to lines in the EIS SW channel by a
factor of up to two. The \ion{Fe}{xxiv} results found here appear to
confirm this.

\subsection{Dynamics: Doppler shifts}\label{sect.vel}

The line-of-sight (LOS) velocities given in Table~\ref{tbl.eis} are derived
from Doppler shifts of the lines relative to ``rest''
wavelengths. There is no direct means to determine an absolute
wavelength scale for EIS, and so indirect methods are needed -- see
\citet{young12} for a discussion.  For the present work, we 
used the 10 Y-pixels at the bottom of the \ion{Fe}{xv} \lam284.16
slit image
to generate a pixel mask that represented a background region. The
region at the bottom of the raster, while still dominated by active
region plasma, is well-separated from the locations where the main flare
dynamics occur, and thus is relatively stable.  From these background
spectra, the centroids of the following lines are measured:
\ion{Fe}{x} \lam184.54, \ion{Fe}{xii} \lam\lam192.39, 195.12,
\ion{Fe}{xiv} \lam\lam264.79, 274.20, \ion{Fe}{xv} 
\lam284.16 and \ion{Fe}{xvi} \lam262.98, which are assumed to
correspond to rest wavelengths. The rest wavelengths at the location
of the flare kernel are then determined by simply applying the known EIS slit
tilt values \citep{2010SoPh..266..209K}. The remaining lines are either not
present in the background spectra or are too weak to be measured
reliably. These lines are paired with the nearest of the measured
background lines, and the ``rest separations'' of the lines are
assumed to be those determined from the rest wavelengths within the
CHIANTI atomic database \citep{dere97,chianti71}. This then allows the rest wavelengths of the
lines not measured in the background spectra to be determined from
those lines that are.
This method of determining rest wavelengths is not as accurate as that
described by \citet{young12}, and we estimate an accuracy of
$\pm$~10~\kms. As can be seen from Table~\ref{tbl.eis}, however,
velocities much larger than this are seen in the flare spectra.

\begin{figure}[h]
\epsscale{1.0}
\plotone{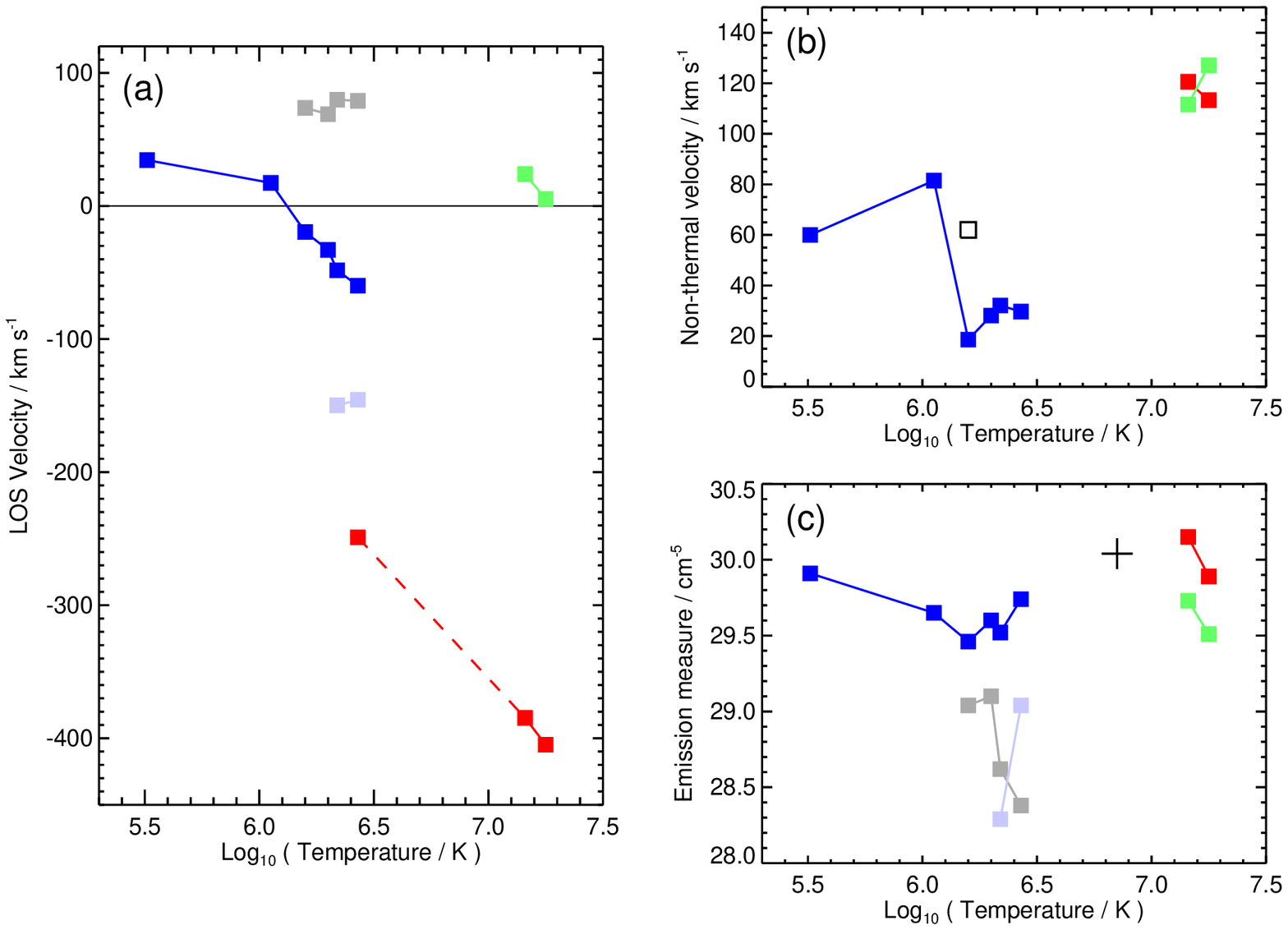}
\caption{Plots showing (a) LOS velocity, (b) non-thermal velocity and
  (c) emission measure values 
derived from the EIS spectrum of the flare kernel. The colors of the points
plotted in (b) and (c) correspond to the groups of velocity components
shown in plot (a). Grey and light blue points are not shown on plot
(b) as they have the same values as the set of blue points. The
unfilled square in plot (b) shows the 
non-thermal velocity derived from \ion{Fe}{xii} \lam192.39 when it is
fit with a single Gaussian.
The cross on plot (c) shows the value
derived from the AIA 94~\AA\ channel.}
\label{fig.vel}
\end{figure}

Figure~\ref{fig.vel}a shows the velocities derived for the EIS
emission lines and their various components. The velocities are
divided into five groups that we believe are physically connected. For
example, the lines of \ion{Fe}{xii--xvi} all show a weak red-shifted
component and are represented by the light gray points. For the lines
formed over $\log\,T=5.5$--6.5, the velocities of the dominant
emission components are represented by the blue points. The change
from redshift to blueshift between \ion{Fe}{x} and \ion{Fe}{xii} is
similar to the patterns found by \citet{milligan09} and
\citet{2010ApJ...719..213W}, which were cited as evidence of explosive
evaporation. A dashed line connects the \ion{Fe}{xvi} high velocity
component to those of \ion{Fe}{xxiii} and \ion{Fe}{xxiv} to indicate
that they might be related, although this is speculative.

The \ion{Fe}{xxiii} and \ion{Fe}{xxiv} lines both show a
highly-blueshifted plasma component that is stronger than the plasma
component that is close to the rest velocity, and the profiles are
shown in 
Fig.~\ref{fig.hot-profiles}. This result demonstrates that a \emph{dominant},
high-velocity upflow component can be seen during flares when a
spectrometer has high spatial resolution. As discussed in the
introduction, earlier X-ray spectra averaged over the whole disk
always revealed a dominant plasma component near the rest velocity of
high temperature lines, with the high velocity component present as a
weak shoulder to this component. A previous observation of a
dominant high velocity component in the \ion{Fe}{xxiii} \lam263.77 line was presented by
\citet{2010ApJ...719..213W} -- see the middle panel of Figure~4 of
their work -- who found a velocity of $-382$~\kms\ from 
during the rise phase of a C9.7 flare. This
suggests that such profiles may be 
typical of M-class flares, although the observation is sensitive to
the timing and position of the EIS slit.

\subsection{Dynamics: non-thermal line broadening}\label{sect.wid}

The measured widths of the EIS emission lines comprise three
components: an instrumental width, a thermal width, and a non-thermal
width. The instrumental width is known and varies with the positions of the
spectra along the EIS slit \citep{eis_sw7}. The thermal width is
determined by assuming an isothermal temperature for the plasma
emitting the particular ion species under consideration. For the
present case we use the $T_{\rm mem}$ values given in Table~\ref{tbl.eis}. The non-thermal
 width is the remaining line width after the 
thermal and instrumental widths have been subtracted and is usually
expressed as a velocity, $\xi$, defined as 
\begin{equation}
4 \ln 2 \left( {\lambda\over c} \right)^2 \xi^2 = W^2 - W_{\rm I}^2 - W_{\rm th}^2,
\end{equation}
where $\lambda$ is the wavelength of the emission line, $c$ the speed
of light, $W$ is the full-width at half-maximum (FWHM) of the fitted
Gaussian function, $W_{\rm I}$ is the instrumental width and $W_{\rm
  th}$ is the thermal width, both expressed as FWHM values. 
The uncertainties on $\xi$ are determined from the 1-$\sigma$
measurement errors
on $W$ and an estimated uncertainty of 3~m\AA\ on $W_{\rm I}$ \citep{eis_sw7}.

The values of $\xi$ are given in Table~\ref{tbl.eis} and displayed
graphically in Figure~\ref{fig.vel}b. The blue points show the
values derived from \ion{O}{vi} through \ion{Fe}{xvi}.
As noted previously, the emission
lines for \ion{Fe}{xii}--\ion{Fe}{xvi} were fit with multiple
Gaussians forced to have the same width, and it is noticeable that a large decrease in $\xi$ occurs
between \ion{Fe}{x} and \ion{Fe}{xii}. We believe this is due to
the fact that the \ion{Fe}{x} line was fit with one Gaussian, while
\ion{Fe}{xii} was fit with two. The black, unfilled square in
Figure~\ref{fig.vel}b shows the non-thermal velocity if the \ion{Fe}{xii} line is
fit with a single Gaussian, giving a much higher value. This suggests
that the large width of the \ion{Fe}{x} line arises from multiple
velocity components that are unresolved in the line profile.

The relatively small non-thermal velocities of \ion{Fe}{xii--xvi}
suggest that, when the background coronal emission is subtracted and
individual plasma components can be identified through multiple
Gaussian fits, then these components do not show any extra broadening
over typical coronal plasma. Therefore enhanced broadening of coronal lines
found in previous measurements is likely to be due to the
superposition of multiple plasma components with different velocities
rather than, say broadening due to turbulence.

The red and green points in Figure~\ref{fig.vel}b show the $\xi$
values for the fast upflowing and stationary components of the
\ion{Fe}{xxiii} and \ion{Fe}{xxiv} lines (see also
Figure~\ref{fig.vel}a), and they are found to be very similar at
around 120~\kms. \citet{milligan11} previously presented non-thermal
velocity measurements from a flare kernel and found remarkably similar
values to those found here, although in that case only an at-rest
plasma component was detected.

\subsection{Emission measure}

The fourth quantity shown in Table~\ref{tbl.eis} is the column
emission measure, which is determined directly from the measured line
intensity through the expression:
\begin{equation}\label{eq.em}
4\pi I= {hc \over \lambda} \epsilon(X) C_\lambda EM(d)
\end{equation}
where $I$ is the measured line intensity, $h$ Planck's constant, $c$
the speed of light, $\lambda$ the wavelength of the emission line,
$\epsilon(X)$ the element abundance of element X relative to hydrogen,
$C_\lambda$ contains various atomic parameters and $EM(d)$ is the
column emission measure.  The atomic parameters contained in
$C_\lambda$ are computed using the CHIANTI routine INTEGRAL\_CALC,
available in \emph{Solarsoft}. The temperatures shown in
Figure~\ref{fig.vel}c are the temperatures of maximum emission, $T_{\rm
  mem}$. The coronal abundances of \citet{2012ApJ...755...33S} have been used, the
ionization balance calculations are those from CHIANTI \citep{dere09},
and a density of $3.4\times 10^{10}$~cm$^{-3}$ was assumed (see Sect.~\ref{sect.dens}).
We note that the emission measure values are insensitive to the precise
density used, however.

An additional emission measure point shown in Figure~\ref{fig.vel}c was
derived from the AIA 94~\AA\ channel as follows. At temperatures of
$\approx$ 10~MK, this channel is dominated by \ion{Fe}{xviii}
\lam93.93 and so an emission measure can be derived by assuming that
the plasma is isothermal at the $\log\,T_{\rm  mem}=6.85$ value of
this line. We compute an isothermal spectrum from CHIANTI for the A94
channel, convolve it with the A94 response function, and sum the
result. The count rate per second in the A94 channel, $C$, (expressed as
data numbers per second) can then be related to the column emission
measure as
\begin{equation}\label{eq.aia-em}
EM = 5.25 \times 10^{26} \alpha C,
\end{equation}
where $\alpha$ accounts for the difference in spatial pixel size
between AIA and EIS due to the fact that the flare kernels are not
resolved by either instrument (see Sect.~\ref{sect.aia}), and takes a
value of 0.18.
The
isothermal spectrum used for this calculation was computed using the
CHIANTI ionization balance, 
the hybrid coronal abundances of \citet{2012ApJ...755...33S}, a density of
$3\times 10^{10}$~cm$^{-3}$, and unit column emission measure; the
pre-flight A94 response function, available through the
\emph{Solarsoft} AIA\_GET\_RESPONSE routine, was used. 

To determine the A94 count rate, the EIS-AIA co-alignment procedure
(Sect.~\ref{sect.coalign}) enabled the position of the EIS slit relative
to AIA to be found. The region in the AIA image corresponding to the
2\arcsec\ wide EIS slit was extracted, the A94 counts summed over the
brightening, and a background level subtracted, leaving a total count
rate of 11,470~DN~s$^{-1}$. The A94 emission measure is then
$6.02\times 10^{30}$~cm$^{-5}$.  Due to the method used we 
consider this an upper limit to the real emission measure.

Figure~\ref{fig.vel}c shows the A94 emission measure point lies within a
temperature gap in the EIS 
emission measure distribution, and thus provides valuable information. EIS does have access to
\ion{Ca}{xvii} \lam192.86 (formed at 6~MK) and unblended
\ion{Fe}{xvii} lines (formed at 4~MK), but these lines were not
obtained in the present study.

Including the AIA point, the emission measure distribution is fairly
uniform from 0.3~MK to 30~MK. The rest components of
the hottest ions (green points in
Figure~\ref{fig.vel}c) have similar values to those derived from the dominant emission
components of the cooler ions (blue points in
Figure~\ref{fig.vel}c). The fast upflowing
hot plasma (red points) has an emission measure around a factor 2.5
larger than that for the rest
components, and the A94 emission measure value is closer to this
value, perhaps suggesting that \ion{Fe}{xviii} also has a dominant
intensity component from the upflowing plasma. The redshifted components of \ion{Fe}{xii--xvi} have
significantly lower emission measures (gray points in
Figure~\ref{fig.vel}c), with a sharp drop between
\ion{Fe}{xiv} and \ion{Fe}{xv}, while the fast upflowing component of
\ion{Fe}{xvi} is much stronger than that of \ion{Fe}{xv} (light blue
points). 

% The emission measure distribution derived from the dominant emission
% components of the cooler ions (blue points in
% Figure~\ref{fig.vel}c) is fairly uniform, and the rest components of
% the hot ions (green points in
% Figure~\ref{fig.vel}c) also have similar values. The fast upflowing
% hot plasma (blue points in
% Figure~\ref{fig.vel}c) has an emission measure around a factor 2.5
% larger than that for the rest
% components. The redshifted components of \ion{Fe}{xii--xvi} have
% significantly lower emission measures (gray points in
% Figure~\ref{fig.vel}c), with a sharp drop between
% \ion{Fe}{xiv} and \ion{Fe}{xv}, while the fast upflowing component of
% \ion{Fe}{xvi} is much stronger than that of \ion{Fe}{xv} (light blue
% points). 

% The AIA emission measure value is almost an order of magnitude larger
% than the values derived from EIS, but this must be considered
% uncertain due to the method used and the possibility of a calibration
% difference between the two instruments. EIS does have access to
% \ion{Ca}{xvii} \lam192.86 (formed at 6~MK) and unblended
% \ion{Fe}{xvii} lines (formed at 4~MK) that would be a check on the A94
% value, but these lines were not obtained in the present study.

All of the EIS emission lines used in the present study are from iron
ions, except for \ion{O}{vi} and we note that the choice of abundance
file would affect the \ion{O}{vi} emission measure value. If
photospheric abundances had been used, then the \ion{O}{vi} emission
measure value would be reduced by a factor three relative to the iron values.

\subsection{Density and emitting volume}\label{sect.dens}

Two density diagnostics are available from the HH\_Flare\_180x160\_v2
study: \ion{Fe}{xii} \lam195.18/\lam195.12 and \ion{Fe}{xiv}
\lam264.79/\lam274.20. The \ion{Fe}{xii} lines
 can not be accurately separated due to the non-Gaussian profiles in
the flare kernel, but  the  \ion{Fe}{xiv} ratio is useful
 and has been applied in previous EIS analyses by, for
example, \citet{doschek07}, \citet{milligan11} and
\citet{2011A&A...526A...1D}. 

Table~\ref{tbl.fe14} shows the electron number densities,
$N_{\rm e}$, derived from the \ion{Fe}{xiv} line intensities given in
Table~\ref{tbl.eis} for the two plasma components. Component 1 has the
stronger emission in both lines and has a velocity of $\approx$
$-30$~\kms, while component 2 has a velocity of 
$+65$ to $+70$~\kms.
% \lam264.79/\lam274.20
% ratios. 
Atomic data for the 
density diagnostic come from version 7.1 of the CHIANTI database
\citep{chianti71}.

\begin{deluxetable}{lccccc}
\tablecaption{Quantities derived from the \ion{Fe}{xiv}
  \lam264.79/\lam274.20 ratio.\label{tbl.fe14}}
\tablehead{
  \colhead{Plasma} &
  \colhead{Ratio} & 
  \colhead{$\log\,(N_{\rm e}/{\rm cm}^{-3})$} &
  \colhead{$d$/arcsec} &
  \colhead{$d_{\rm c}$/arcsec} \\
  \colhead{Component} & 
  }
\startdata
1 & $2.44\pm 0.05$ & $10.53\pm 0.05$ & $4.77\pm 1.10$ & 2.12 \\
2 & $2.02\pm 0.14$ & $10.15\pm 0.11$ & $8.44\pm 4.30$ & 2.57  \\
\enddata
\end{deluxetable}

The column depth, $d$, is an important parameter that can be derived from
spectroscopic data as it can potentially yield smaller length scales
than those obtained by directly imaging a plasma. For example,
\citet{2011A&A...526A...1D} found a column depth of 10~km for a flare
kernel observed by EIS. Unfortunately the column depth also has
large uncertainties associated with it due to the dependence on the
square of the
density -- the uncertainties shown in Table~\ref{tbl.fe14} are derived
by propagating the uncertainties on $I$ and $\log\,N_{\rm e}$. In
addition, we cannot be certain of the absolute iron abundance: the
\citet{2012ApJ...755...33S} hybrid coronal abundance value\footnote{We
  use the standard abundance notation here whereby the abundance is
  expressed as $\log\,\epsilon({\rm
    Fe})+12$.}  of
7.85  lies
between the \citet{feldman92} coronal value of 8.10 that is commonly used, and the
photospheric value of 7.52 \citep{2011SoPh..268..255C}, thus
introducing an
additional 
factor two uncertainty.

A further uncertainty lies in the EIS absolute
calibration. Pre-launch, the uncertainty was considered to be 22\%\ 
\citep{2006ApOpt..45.8689L} but the degree to which the instrument
sensitivity has degraded, and any wavelength sensitivity that this
degradation has, are uncertain at the time of writing. As
highlighted by the \ion{Fe}{xxiv} \lam192.03/\lam255.11 ratio
discussed earlier, a factor of two uncertainty is possible. 

The values of $d$ derived here  may thus, at worst, be
uncertain to a factor four, however the relative values of $d$ between the two
plasma components is much more accurate as the abundance and absolute
calibration uncertainties do not apply. Implicit in the calculation of
$d$ is the assumption that the emission line intensities come from a
region with cross-sectional area 2\arcsec\ $\times$ 1\arcsec. If we
assume instead that the emitting volume is a cube, then the side of
this cube is $d_{\rm c}=\sqrt[3]{2d}$ -- a parameter that we refer to
as the cubic column depth. We discuss the size of the emitting volume
in more detail in the following section, where higher resolution AIA
images are presented.

\section{SDO data analysis}\label{sect.aia}

\begin{figure}[h]
\epsscale{1.0}
\plotone{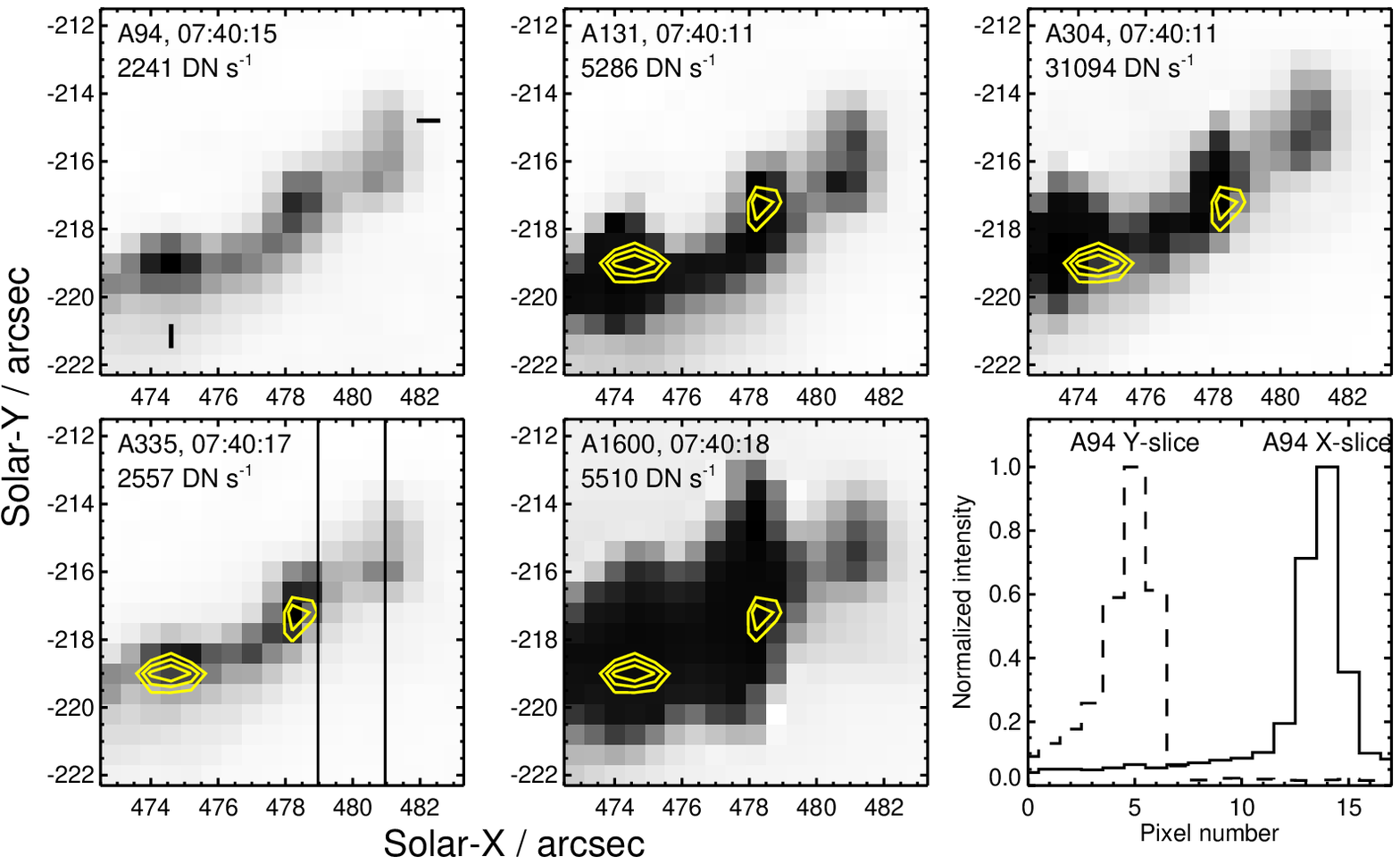}
\caption{The top three panels, and two lower-left panels show AIA images
  of the flare kernel site at times ranging from 07:40:11 to
  07:40:18~UT. The intensity scale is reversed such that dark areas
  correspond to high intensity. For each image the count rate of the brightest pixel in
the image is displayed in the top-left. The yellow contours show the
areas of highest
intensity from the A94 image. The two vertical lines on the A335 image
show the position of the EIS slit as determined from the co-alignment
method. The two short, thick lines in the A94 image indicate the X and
Y pixels used to create the intensity cross-sections plotted in the
bottom-right panel.}
\label{fig.aia-ims}
\end{figure}

Figure~\ref{fig.aia-ims} shows images of the flare kernel site
obtained by the AIA instrument. The mid-point of the EIS exposure that
produced the kernel spectrum discussed in the previous section was 
07:40:16~UT and so the AIA images closest to this time were
chosen. AIA has nine different EUV and UV filters, but four of these
gave badly saturated images. Even for the images displayed, only A94
and A335 are not saturated at any location.

An important point to note is that the basic morphology of these
images is very similar, and inspection of other data from the sequence
of AIA images confirms that this is generally true. This is also
consistent with the intensity images from EIS
(Figures~\ref{fig.eis-ims} and \ref{fig.eis-xs}). We are thus
confident that the flare kernels emit over a continuous range of
temperatures from the chromosphere (as observed through the 1600 and
1700~\AA\ filters of AIA) through to temperatures of $\approx$~30~MK
(the \ion{Fe}{xxiv} emission lines observed by EIS).

The yellow contours from the A94 image suggest that there are slight
spatial offsets between the different AIA filters of up to 2
pixels. These are consistent with the expected accuracy of the AIA
image alignments (R.A.~Shine, private communication 2011), and so it
is likely that the flare kernels are co-spatial at different temperatures.

The size of the flare kernels is very small and in fact the AIA
de-spiking routine sometimes flags them as cosmic rays. For this
reason it is necessary to use the AIA\_RESPIKE \emph{Solarsoft} routine
to put flagged spikes back into the data. Two A94 intensity cross-sections
are shown in the lower-right panel of Figure~\ref{fig.aia-ims}; one in
the X-direction and one in the Y-direction. The column and row chosen for
these cross-sections are indicated on the A94 image. The two different
flare kernels shown  have narrow, Gaussian shapes and by selecting
a few such cross-sections we find average FWHMs of 2.2 and 2.0 pixels
for the Y and X directions, respectively, with an uncertainty of about
0.10 pixels. These values are actually narrower than the AIA spatial
resolutions of 2.5--3.0~pixels quoted in \citet{boerner12}, 
 suggesting
the instrument is performing somewhat better than expected.

The twin vertical lines over-plotted on the A335 image of
Figure~\ref{fig.aia-ims} show the location of the EIS slit as
determined from the co-alignment method described in
Appendix~\ref{sect.coalign}. It can be seen that EIS does not
necessarily observe a single flare kernel, but instead a patch that
may include two or more. This complicates interpretation of the column
depth values $d$ and $d_c$ discussed in Sect.~\ref{sect.dens}. The
narrow intensity width across the line of flare kernels suggests an actual
width significantly smaller than an AIA pixel. We can thus make a
simple model whereby the emission is uniform along the line of kernels
in the region observed by the EIS slit, and has a width across the
kernel line of 0.3\arcsec\ (half of an AIA pixel). This would then imply the column depth of
the kernel site is $d/0.3= 16$\arcsec\ from the EIS \ion{Fe}{xiv}
diagnostic. However, the appearance of 
the kernels in
the AIA image should then be a series of ``spikes'', unless the spikes
are aligned along the observer's line of sight. We conclude that the
column depth derived from EIS is thus incompatible with the kernel sizes
observed by AIA. The discrepancy could be resolved through one or more
of the following: (i) the density is actually higher than derived from
the \ion{Fe}{xiv} ratio, (ii) the EIS sensitivity is higher than
assumed at the wavelengths of the \ion{Fe}{xiv} lines, and (iii) the
abundance of iron is higher than assumed.

\begin{figure}[h]
\epsscale{0.7}
\plotone{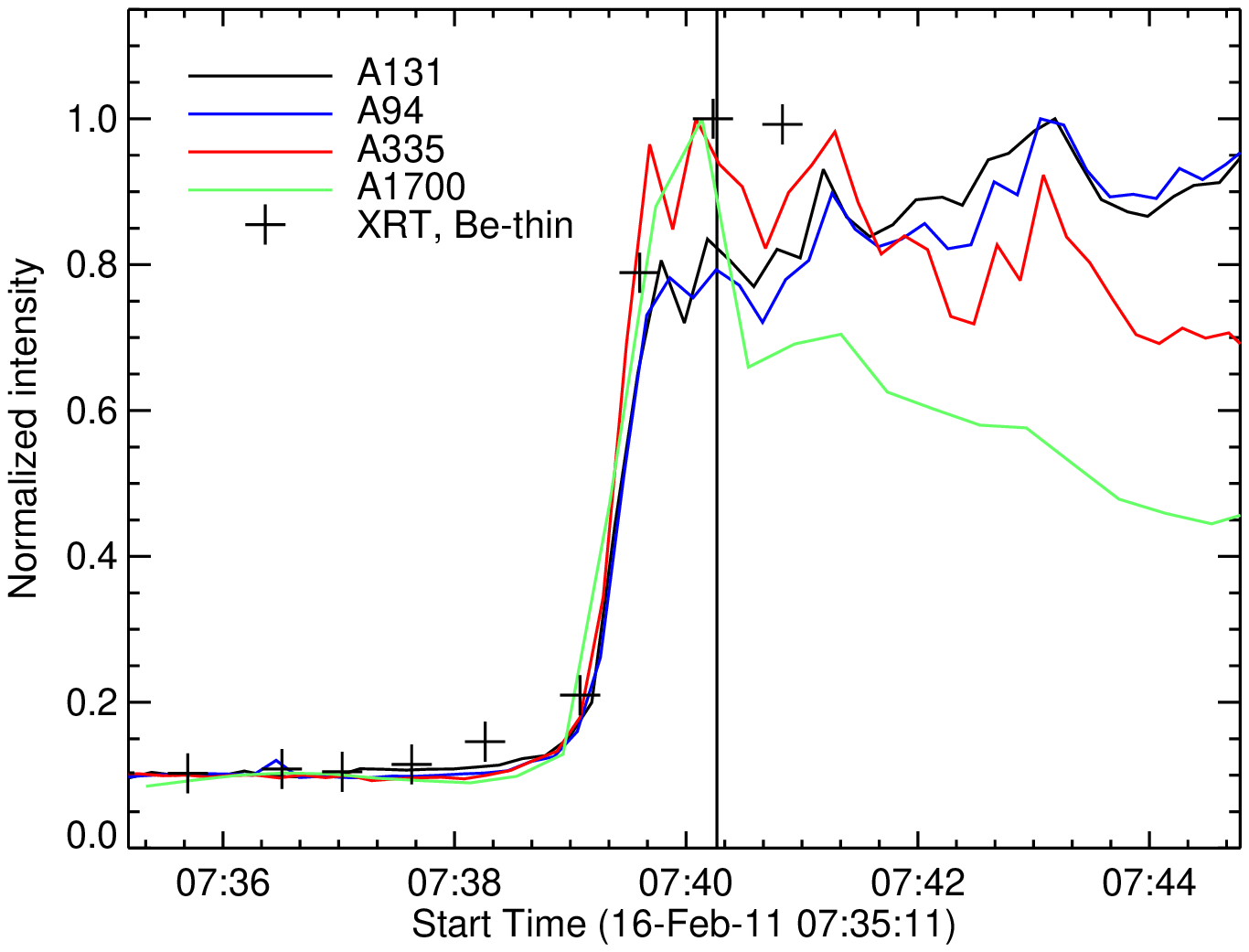}
\caption{AIA light curves for the EIS flare kernel. Each curve has
  been normalized such that the median of the values before 07:38 is
  set to 0.1 and the maximum of the curve is set to 1.0. The vertical
  line denotes the midpoint time of the EIS exposure. A XRT light
  curve derived from the Be-thin filter is over-plotted with crosses.}
\label{fig.aia-lc}
\end{figure}

With the approximate position of the EIS slit established, it is 
possible to construct light curves for the kernel  site observed by
EIS in different AIA
filters, and Figure~\ref{fig.aia-lc} shows four such curves (the
remaining AIA filters are affected by saturation). The light curves
were derived by taking AIA images at the highest cadence (12~s for the
EUV filters; 24~s for A1700) and extracting the spatial region
corresponding to the 2\arcsec\ $\times$ 9\arcsec\ region that was used
to create the EIS spectrum. The counts in this region were then summed.
For Figure~\ref{fig.aia-lc} the four light curves were normalized such
that they each have the same pre-flare intensity and the same maximum
intensity. It is clear that all four channels brighten simultaneously
(at least within the resolution of the AIA instrument), with the rise
to maximum taking place in about 40~s from 07:39:00 to
07:39:40~UT. There is also a small intensity rise (a few percent) in all
channels beginning at 07:38:00~UT which may be related to the
following large
increase. The EIS observation begins within 30~s of the kernel
reaching its maximum intensity level.

\begin{figure}[h]
\epsscale{0.5}
\plotone{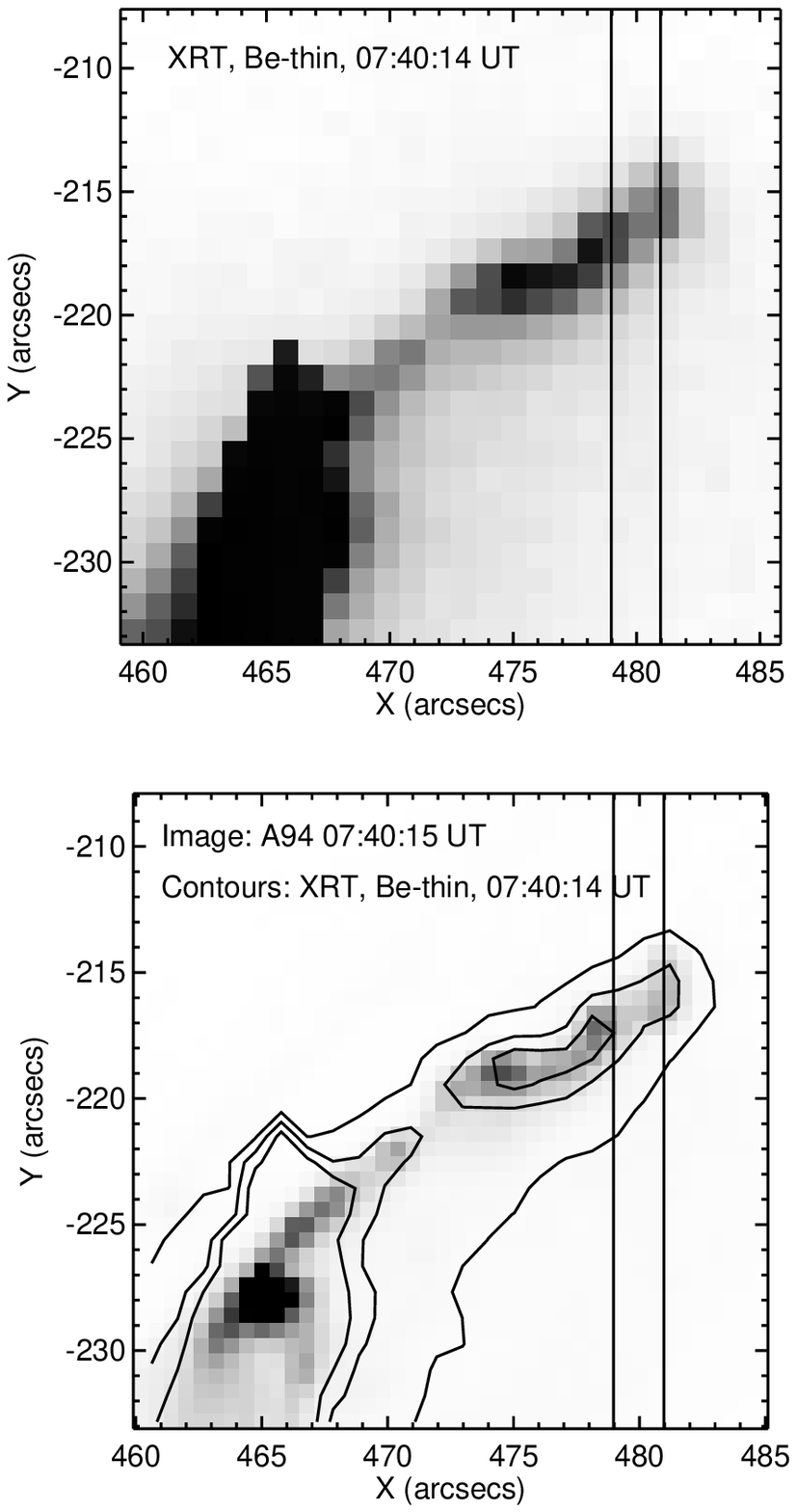}
\caption{The upper panel shows an XRT image of the flare brightenings obtained at 07:40:14~UT
  with the thin beryllium filter. The lower panel shows the A94 image
  from 07:40:15~UT. The twin vertical lines show the location of the
  EIS slit at 07:40:16~UT. The contours on the lower panel show the
  XRT emission. }
\label{fig.xrt}
\end{figure}

A further light curve is shown in Figure~\ref{fig.aia-lc} and was
obtained from the X-Ray Telescope (XRT) on board \hinode. We have
chosen data from the thin beryllium filter (Be-thin) which were
obtained at about 30--45~s cadence over the period 07:35 to
07:41~UT. The observing mode changed after 07:41~UT, presumably because
a flare mode was triggered, and Be-thin images were not obtained again
until 07:50~UT.  The flare kernels are prominent in the Be-thin images
and Figure~\ref{fig.xrt} compares the 07:40:14~UT image with the A94
image from 07:40:15~UT (see also Figure~\ref{fig.aia-ims}). The XRT
image has been co-aligned with the A94 image using the brightenings in
the middle-right of the images. Note that part of the XRT image is
saturated. Based on the displayed co-alignment, the XRT light curve
was extracted and normalized in the same manner as for the AIA images
discussed earlier. The Be-thin filter has a peak temperature response at
$\log\,T=7.0$ \citep{2007SoPh..243...63G} and confirms the high
temperatures present in the flare kernel.

At this point we note that \citet{2012A&A...540A..24B} recently
suggested that all of the AIA channels can 
brighten simultaneously in the early stages of a flare due to a large
increase in emission at temperatures of 0.1--0.7~MK. This is evidenced
by the fact that all of the AIA channels brightened simultaneously
and several minutes before the hard X-ray bursts for a B5 microflare
observed on 2010 July 21, in a manner consistent with previous
observations of transition region emission lines observed with the
SOHO/CDS instrument. The XRT light curve demonstrates that this is not
the case in the present flare as the Be-thin filter has negligible
sensitivity below $\log\,T=6.0$ \citep{2007SoPh..243...63G} while, in
addition, the EIS emission measure plot of Figure~\ref{fig.vel}c
shows that the emission measure at $\log\,T=5.5$ is not significantly
larger than at other temperatures observed by EIS. The flare kernel
studied here can thus not be explained as a 
transition region event.

\section{Relation to magnetic field}\label{sect.mag}

\begin{figure}[h]
\epsscale{0.5}
\plotone{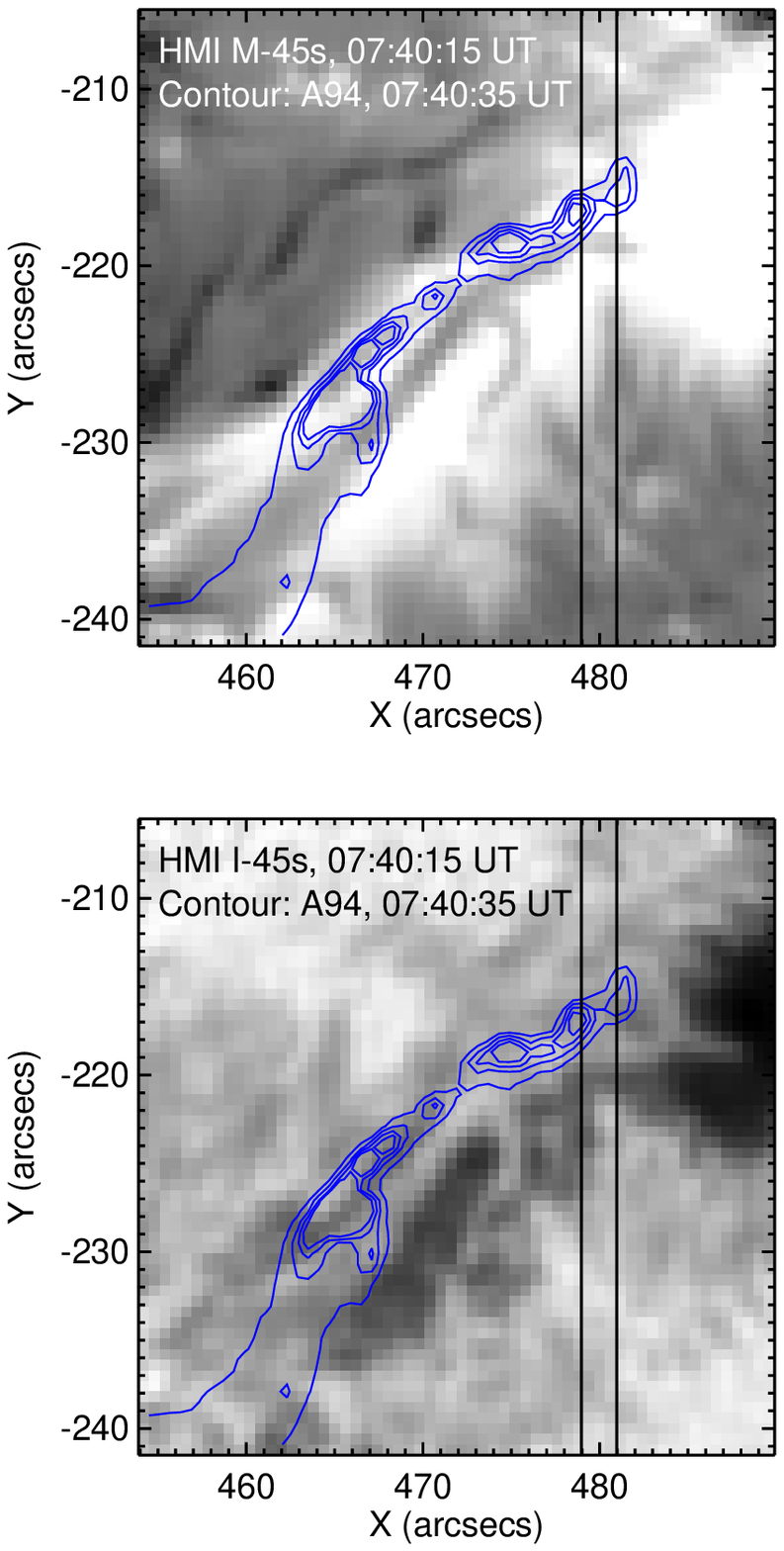}
\caption{The upper panel shows a HMI LOS magnetogram, and the lower
  panel a HMI white light continuum image. The blue lines show
  intensity contours from the AIA 94~\AA\ image. The parallel vertical
lines indicate the position of the EIS slit at 07:40:16~UT.}
\label{fig.hmi-0740}
\end{figure}

Figure~\ref{fig.hmi-0740} shows where the flare kernels from the A94 07:40:15~UT
exposure occur in relation to the LOS magnetic field and visible
continuum intensity as measured from the
HMI instrument. Alignment between the two instruments was checked by
comparing images obtained 150\arcsec\  to solar-north and south of the
region displayed in Figure~\ref{fig.hmi-0740}. In such quiet Sun
regions, small bright points in  A1700 images generally correspond
well with bright points seen in plots of the absolute LOS  magnetic
field strength. In the present case it was found that changing the
A1700 image center by $(+0.3\arcsec ,-0.5\arcsec)$ gave an improved
alignment between the bright points for both quiet Sun pointings.

Now, comparisons of  A1700 and A94 images of the flare kernels suggest
that there is a spatial offset of 
$(0.0\arcsec ,-0.8\arcsec )$ between the two, i.e., the A94 image needs to be moved to
solar-north to better match the A1700 image. Therefore the offset
between A94 and the HMI LOS  magnetograms is $(+0.3\arcsec
,+0.3\arcsec)$.  Given the assumptions made in this estimate, the
accuracy may be as large as $\approx$1\arcsec, however we can
have some confidence that the brightenings are aligned along the ridge
of positive magnetic polarity in Figure~\ref{fig.hmi-0740} (upper panel),
rather than the narrow channel of weak opposite polarity just to the
south of 
it, or the area of weak magnetic field to the north. Averaging a
number of spatial pixels around the location of the flare kernel site
considered in the present work yields an average line-of-sight
magnetic field strength of $1030\pm 150$~G. The comparison
with the continuum intensity image (lower panel of
Figure~\ref{fig.hmi-0740}) shows that the brightenings lie within
penumbra regions and extend towards, but not into, the sunspot umbra
on the right side of Figure~\ref{fig.hmi-0740}.

The HMI magnetograms are obtained at 45~s cadence and it is possible
to search for changes in the signal between frames. Changes of up to
100~G are found at the location of one of the kernels between
frames at 07:39:05 and 07:39:50~UT, and between 07:40:35 and and
07:41:20~UT, however it is likely that such changes may simply represent
plasma dynamics (velocity shift and/or line broadening) in the
\ion{Fe}{i} line used for the magnetogram measurements rather than an
actual magnetic field change, and so we choose not to identify them as
magnetic field changes.

\section{Discussion and summary}\label{sect.summary}

Active region AR 11458 produced a confined M1.1 class flare on 2011 February 16
that peaked at 07:44~UT. On one side of the active region a number of
intense flare kernels were observed spectroscopically by the \hinode/EIS instrument
between 07:40 and 07:41~UT. The present work focussed on one flare
kernel site observed by EIS at 07:40:16~UT, and various spectroscopic
parameters were measured. In addition, images from the SDO/AIA
instrument were used to study temporal and spatial properties of the
kernel site. We believe the analysis presented here gives the most complete
set of ultraviolet observations yet obtained of flare kernels and
present an important reference data-set against which other
observations of these important energy release events can be
compared. The key results are summarized below.

AIA and XRT images demonstrated that the flare kernel site reached
maximum brightness in about 40~s, with a weak intensity enhancement
for a minute immediately prior to this. Of the four AIA channels that
were not saturated (spanning temperatures from the chromosphere to
10~MK) the intensity increase was simultaneous within the resolution
limits of the AIA instrument. 
The flare kernel  is one of several that
lie along a line of length $\approx$~25\arcsec\ that is found to align
with a ridge of strong, positive magnetic field that is related to a
nearby sunspot. The kernel is  located in the penumbra of this
sunspot where the magnetic field strength is $\approx 1000$~G.
The flare kernel sizes are at the
resolution limit of the AIA instrument, suggesting sizes of
$<0.6$\arcsec\ ($\lesssim 400$~km). In addition they have a similar
morphology at all temperatures from the chromosphere to 30~MK, and are
co-spatial to within the alignment uncertainties of the AIA channels.

A single EIS spectrum of the flare kernel was obtained about 30~s
after the kernel reached maximum intensity, and the following
properties were determined.

\begin{itemize}
\item The LOS velocities of the dominant emission components of lines from \ion{O}{vi} to
  \ion{Fe}{xvi} (0.3--2.5~MK) decrease monotonically  from $+35$~\kms\
  (downflows) at $\log\,T=5.5$ to $-60$~\kms\ (upflows) at
  $\log\,T=6.4$. The transition from downflows to upflows occurs around
  $\log\,T=6.1$.
\item \ion{Fe}{xxiii} and \ion{Fe}{xxiv} lines (formed at 10--30~MK)
  show two plasma components, one at $\approx$~-400~\kms, and a
  weaker one at $\approx$~0~\kms.
\item Lines from \ion{Fe}{xii--xvi} show two to four emission
  components. Each line has a red-shifted component at around
  $+60$ to $+70$~\kms, while \ion{Fe}{xv} and \ion{Fe}{xvi} have
  blue-shifted components at around $-150$~\kms. There is also
  evidence for a further \ion{Fe}{xvi} component at $-250$~\kms.
\item All lines show non-thermal velocity broadening, although for
  \ion{Fe}{xii--xvi} -- lines possessing multiple plasma components --
  the
  values are quite small at around 20--30~\kms. The cooler \ion{O}{vi}
  and \ion{Fe}{x} lines have larger broadenings of 60--80~\kms,
  perhaps reflecting unresolved plasma components. The hottest lines,
  \ion{Fe}{xxiii} and \ion{Fe}{xxiv} have large broadenings of around
  100--120~\kms.
\item Emission measure values derived from EIS are fairly uniform with
  temperature, 
  with values of 3--10 $\times 10^{29}$~cm$^{-5}$. The AIA 94~\AA\
  emission measure value fills a gap in the temperature coverage of
  EIS and is consistent with the EIS values.
\item The dominant, blueshifted emission component of \ion{Fe}{xiv}
  (2~MK) has a density of $3.4\times 10^{10}$~cm$^{-3}$; the weaker,
  redshifted component has a density of $1.4\times
  10^{10}$~cm$^{-3}$. The column depths implied by these densities are
  significantly larger than the observed sizes of the flare kernels.
\end{itemize}

Prior to the availability of AIA and EIS, the limited spatial
resolution of previous flare kernel observations 
prevented direct comparisons with theoretical models of the
chromospheric evaporation process, and so one method for modeling the
solar emission lines was to 
construct ensemble models of multiple events to match the
data
\citep{1997ApJ...489..426H,1998ApJ...500..492H,2005ApJ...618L.157W}. The
new data from AIA and \hinode\ presented here demonstrate 
that it is possible to obtain high quality data of individual
kernel sites. Many 1D models of the evaporation process have been
performed previously and we briefly consider the model of 
\citet{2005ApJ...630..573A}, which is a development of the earlier work
of \citet{1985ApJ...289..414F}, \citet{1994ApJ...426..387H} and
\citet{1999ApJ...521..906A}. 

Firstly, the observation of
upflowing plasma at 400~\kms\ is clear evidence for explosive
evaporation and by comparing with the light curves from XRT and AIA we
can say that chromospheric evaporation was underway 80~s after line
intensities began their large increase, and about 30~s after the
intensities reached their peak value. For comparison, the models of
\citet{2005ApJ...630..573A}  showed that explosive evaporation began
73~s and 1~s after the start of events with heat fluxes of $10^{10}$
and $10^{11}$~\ecs, respectively. The time taken to reach maximum
intensity is around 80~s and 2~s \citep[based on Figure~15
of][]{2005ApJ...630..573A} for these two cases compared to the observed
rise of 40~s. We speculate that the rise time observed from AIA data
may serve as a proxy for the heat flux, and thus the heat flux for the
present flare kernel may be between $10^{10}$
and $10^{11}$~\ecs.

The small size of the flare kernels as seen by AIA suggest scales of
$<0.6$\arcsec\ ($\lesssim 0.4$~Mm) at all temperatures.
The 1D models of
\citet{2005ApJ...630..573A} were computed over heights of 0 to 10~Mm,
with chromospheric lines formed over 0--1.5~Mm, and coronal lines over
1--10~Mm. Taking into account density, the coronal emission is likely
concentrated over 1--3~Mm regions. The observations are thus not
inconsistent with the models. The high densities measured from the EIS
\ion{Fe}{xiv} diagnostic are consistent with the later phases of the
\citet{2005ApJ...630..573A} models when dense chromospheric plasma has
been heated to coronal temperatures.

\citet{2005ApJ...630..573A} focussed on the modeling of chromospheric
emission lines, and so comparisons with the higher temperature EIS
velocities are not possible. We note that the measurement of multiple
emission components of the lines from \ion{Fe}{xii} to \ion{Fe}{xvi},
together with the large line widths of other lines,
suggest that there are several flow patterns within the flare kernel,
which may simply imply that there are multiple loop footpoints within the
kernel that are not resolved by EIS. If this is the case, though, then there is some coherence
between how these footpoints behave as evidenced by distinct velocity
features in the line profiles.

Finally, the emission measure results show fairly uniform emission
over the temperature range 0.3--30~MK, which constrains the heating
profile for upper transition region and coronal plasma.

The similarity of the flare kernel results presented here with those
of \citet{2010ApJ...719..213W} for a similar size flare, and also to
aspects of the results from \citet{milligan09},
\citet{milligan11}, \citet{2011A&A...526A...1D} and \citet{doschek12},
suggests that flare kernels may exhibit consistent properties. A
survey of such events with EIS and AIA would thus be extremely
valuable. RHESSI observations are extremely important too as they can
constrain the energy input to the flare kernel site.

\acknowledgments

This work was funded by NASA under a contract to the U.S.\ Naval Research
Laboratory. 
Hinode is a Japanese mission developed and launched by
ISAS/JAXA, with NAOJ as domestic partner and NASA and
STFC (UK) as international partners. It is operated by
these agencies in co-operation with ESA and NSC (Norway).

{\it Facilities:} \facility{Hinode(EIS)}, \facility{Hinode(XRT)},
\facility{SDO(AIA)}, \facility{SDO(HMI)}, \facility{GOES}.

\appendix

\section{Co-aligning EIS and AIA}\label{sect.coalign}

The EIS and AIA data were spatially coaligned by constructing a pseudo
raster image from the AIA 335~\AA\ filter images and comparing with
the EIS raster image obtained in the \ion{Fe}{xvi} \lam262.98
line. The A335 filter is dominated by \ion{Fe}{xvi} \lam335.40 and
\ion{Fe}{xvi} \lam262.98 is a 
strong, clean line in the EIS raster.

Figure~\ref{fig.eis-ims} shows that the flare kernels appeared as
compact brightenings in five adjacent EIS exposures, and the alignment
was performed by trying to reproduce the intensity ratios of these
five brightenings in the A335 pseudo raster image. The A335 images
nearest in time to each of the EIS 
exposures were identified and for each data columns were extracted
corresponding to the expected position of the EIS slit. The EIS slit
width is 2\arcsec, spanning more than three AIA pixels, and partial
intensity columns were considered where necessary to ensure that exactly
2\arcsec\ strips were extracted. The solar-X positions of the EIS slit
were obtained through IDL procedures, and were assumed to have
an offset from the real solar-X positions.

The A335 intensity corresponding to each of the EIS brightenings is
obtained by summing 15 pixels on each side of the intensity maximum in
the simulated intensity column. By adjusting the EIS--AIA offset we
attempted to reproduce the intensity ratios of the five EIS
brightenings.  Simply by inspecting the ratios by eye the appropriate
X-offset could be established to be between $+11$ and
$+12$~arcsec. The final value of $+11.7$~arcsec was determined by
performing a least-squares analysis. 

\begin{figure}[h]
\epsscale{0.9}
\plotone{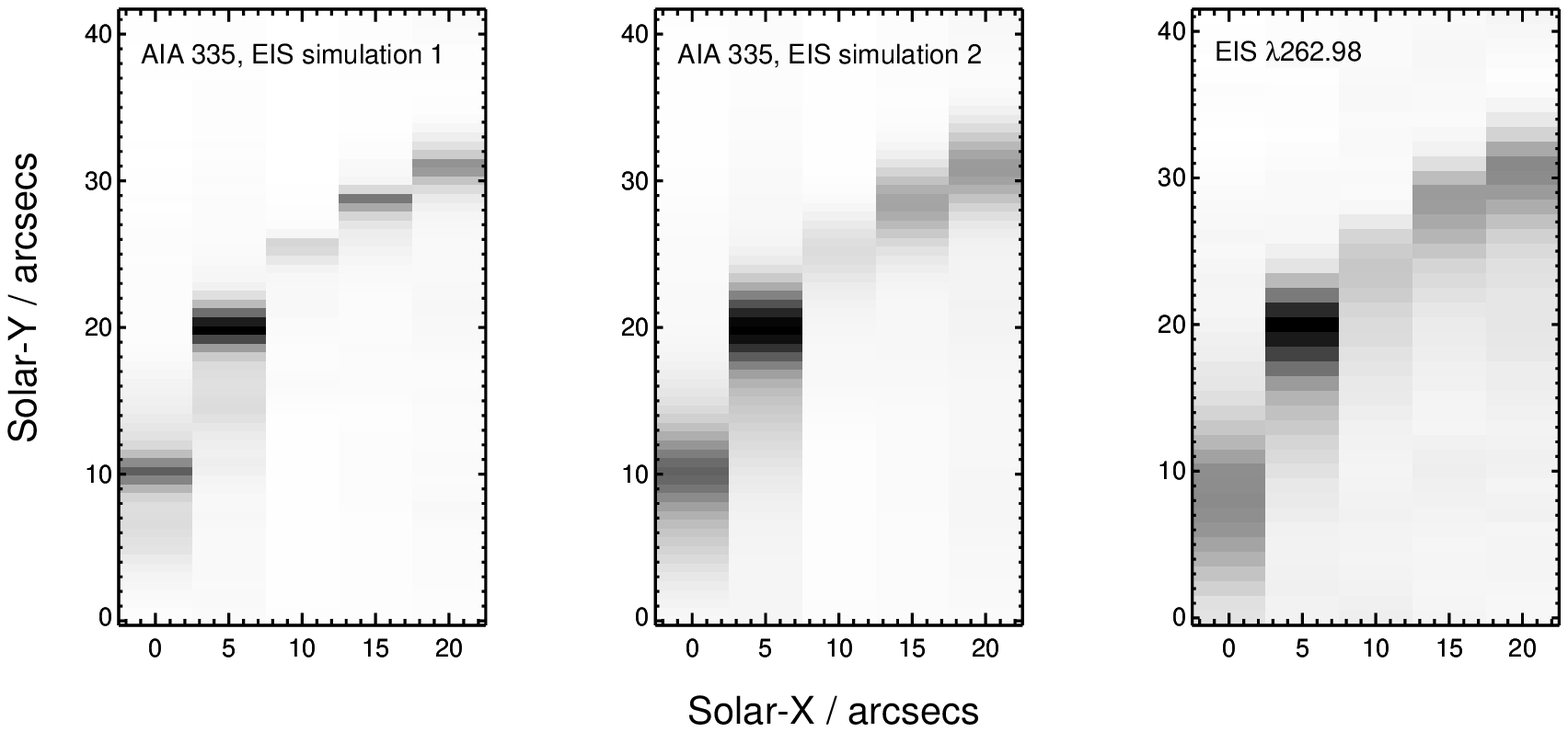}
\caption{Images of the set of five flare brightenings observed by EIS between 07:40 and 07:41~UT. The left image is derived from A335 exposures, where data columns have been summed to match the 2~arcsec\ width of the EIS slit. The middle panel is derived from the left panel image by convolving the data in the Y-direction with a Gaussian of FWHM equal to 3.5~arcsec. The right panel shows the EIS \ion{Fe}{xvi} \lam262.98 intensity image. Note that the displayed pixels have a width of 5~arcsec because the EIS raster used jumps of 5~arcsec between exposures.}
\label{fig.aia-eis}
\end{figure}

 Figure~\ref{fig.aia-eis} (left panel) shows the A335 pseudo raster image derived
 with this method using the 11.7~arcsec offset. This can be compared to
 the actual EIS image in the right panel (this image was derived by
 fitting the \lam262.98 emission line with a Gaussian at each spatial pixel). In
 the middle panel the AIA image has been convolved with a Gaussian of
 FWHM 3.5~arcsec in the Y-direction, to better match the spatial
 resolution of EIS. It can be seen that the correspondence between the
 AIA and EIS images is very good, giving confidence that the
 co-alignment for the brightenings is accurate.

One should caution that the EIS exposure times are significantly
longer than those of AIA (8~s vs.\ 2.9~s), and  there is not a
one-to-one correspondence between EIS and AIA exposures. For such
dynamic events as the flare brightenings this leads to a significant
source of uncertainty in the alignment method presented here, but the
results suggest that an accuracy of 1~arcsec is not unreasonable. 

\bibliographystyle{apj}
\bibliography{myrefs}

\end{document}